\documentclass{article}

\usepackage[preprint, nonatbib]{neurips_2026}

\usepackage[utf8]{inputenc} 
\usepackage[T1]{fontenc}    
\usepackage{hyperref}       
\usepackage{url}            
\usepackage{booktabs}       
\usepackage{amsfonts}       
\usepackage{nicefrac}       
\usepackage{microtype}      
\usepackage{xcolor}         
\usepackage{amsmath,amssymb,amsfonts, amsthm}
\usepackage{algorithm}
\usepackage{algpseudocode}
\usepackage{graphicx}
\usepackage{textcomp}
\usepackage{pgfplots}
\usepackage{tikz}
\usepackage{comment}
\usepackage{subcaption}
\usepackage{tikz}
\usepgfplotslibrary{groupplots}
\usepackage{pgfplots}
\pgfplotsset{compat=1.18}
\usepackage{pgfplotstable}
\usetikzlibrary{pgfplots.fillbetween}

\DeclareMathOperator*{\argmax}{arg\,max}
\DeclareMathOperator*{\argmin}{arg\,min}

\newtheorem{theorem}{Theorem}
\newtheorem{definition}{Definition}
\newtheorem{corollary}{Corollary}
\newtheorem{lemma}{Lemma}

\title{Publishing Below-Threshold Triangle Counts under Local Weight Differential Privacy}

\author{%
  Kevin Pfisterer \\
  The University of Tokyo\\
  Tokyo, Japan \\
  \texttt{pfisterer@is.s.u-tokyo.ac.jp} \\
  \And
  Quentin Hillebrand \\
  University of Copenhagen \\
  Copenhagen, Denmark \\
  \texttt{quentin.hillebrand@di.ku.dk} \\
  \And
  Vorapong Suppakitpaisarn \\
  The University of Tokyo \\
  Tokyo, Japan \\
  \texttt{vorapong@is.s.u-tokyo.ac.jp} \\
}

\begin{document}

\maketitle

\begin{abstract}
 We propose an algorithm for counting below-threshold triangles in weighted graphs under local weight differential privacy. Although many prior studies have considered the setting in which the graph topology is public and only the edge weights are sensitive, to the best of our knowledge, this is the first work to study this privacy notion in the local model. Building on a two-round protocol for locally differentially private triangle counting, we exploit the public graph topology to design a novel algorithmic framework. This leads to significant improvements in both accuracy and scalability. 
 In particular, when the input graph is planar, our algorithm  eliminates the covariance arising from distributed triangle counting at nodes; for graphs with bounded degeneracy, it significantly reduces this covariance. Since covariance is the dominant source of error in the counting task, our method achieves accuracy that closely aligns with the lower bound.
 We also present an efficient algorithm for the computation of the smooth sensitivity and provide experiments that quantify the trade-off between the biased and unbiased variants of our estimator and demonstrate the effectiveness of the proposed improvements.
\end{abstract}

\section{Introduction}
Over recent years, differential privacy \cite{dwork} has emerged as the de facto standard for formal privacy guarantees. It provides a rigorous mathematical framework under which the inclusion or exclusion of any single individual’s data has only a limited effect on the released statistics, thereby preventing an adversary from drawing reliable inferences about any particular data record from the published output.
Initially developed for tabular datasets, differential privacy was subsequently adapted to graph-structured data via edge-differential privacy, which protects the presence or absence of individual edges in a network \cite{nissimsmooth,hay2009accurate}. A key limitation of both the central differential privacy and central edge-differential privacy models is their reliance on a trusted curator to collect and sanitize the data. Local differential privacy \cite{kasiviswanathan_what_2011,qin2017generating} and its graph analogue, local edge-differential privacy \cite{qin2017generating}, address this concern by shifting randomization to the user side, thereby eliminating the need for a trusted central server and providing strong privacy guarantees.

A substantial body of work has studied differential privacy for weighted graphs, which model practical settings such as road and telecommunication networks. In such settings, the graph topology is typically public, and hence the presence or absence of edges is not regarded as sensitive. The sensitive information instead lies in the edge weights, which must be protected. Representative results include differentially private shortest-path algorithms~\cite{sealfon2016,bodwin_2024,chen2023,dinitz2025}, mechanisms for releasing privatized versions of weighted graphs~\cite{liu2024optimal,zou2025almost}, and algorithms for privately releasing cut values~\cite{aamand2024differentially,chandra2024differentially}.

In this work, we strengthen the line of study discussed in the previous paragraph by considering its local analogue. Although several works have studied local differential privacy for weighted graphs~\cite{xu2025continuous,brito2023global,zhu2024local}, to the best of our knowledge, all prior work treats both edge existence and edge weights as sensitive attributes. However, hiding the graph topology is overly restrictive and introduces unnecessary error in applications where the underlying infrastructure is already public knowledge.

The local weight differential privacy framework is well suited to several practical applications. For example, consider a web-traffic model in which nodes represent websites, edges represent hyperlinks, and edge weights capture the volume of users navigating along these links. While the hyperlink structure of the network is public knowledge, the exact traffic volume on each link is sensitive. Since each website can observe referral sources and outbound clicks, it has access to the weights of its incident edges. The local weight differential privacy model therefore enables websites to collaboratively compute global graph statistics while each website independently provides strong privacy guarantees for its own users.

Triangle counting problem has been extensively studied in the unweighted setting under local differential privacy \cite{eden2025triangle, hillebrand_et_al:LIPIcs.STACS.2025.49, imola_2step, signed_triangles, vldb2025}. In the unweighted case, triangle counting plays a central role in applications in graph mining, bioinformatics, and social networks~\cite{newman2009random, assadi_et_al:LIPIcs.ITCS.2019.6, milo2002network}. Since the network topology is a public information in our setting, counting the number of triangles is straightforward in our privacy notion.

\subsection{Our contributions}
We study the problem of counting triangles whose total edge weight is smaller than a given threshold under local weight differential privacy.
This triangle count and related measures such as the weighted clustering coefficient have been used to identify congested regions, which are central to urban traffic planning \cite{qin2020building,saramaki2007generalizations}.
By designing mechanisms that count triangles whose weight lies below (equivalently, above) a given threshold, our approach directly supports this line of work while providing strong privacy guarantees.
Moreover, efficient triangle-counting techniques often extend to more general subgraph counting, as has been shown in the unweighted setting~\cite{betzer2024publishing,suppakitpaisarn2025counting}.

The number of triangles with small or large weights can be used to generate a synthetic graph that preserves key properties of the original graph. Finally, prior differentially private work on shortest-path cycles~\cite{sealfon2016,bodwin_2024,chen2023,dinitz2025} has primarily assumed non-negative edge weights.
Below-threshold triangle identification can also be viewed as a basic building block for detecting negative cycles, with negative triangles being the simplest such structures.
Since negative-cycle detection is fundamental to shortest-path computation with arbitrary edge weights, our results provide an entry point for studying negative-cycle detection under local differential privacy.

In Section \ref{sec:algo}, we propose an algorithm for counting the number of triangles whose total weight falls below a given threshold, fully leveraging the flexibility of the differential privacy notion for weighted graphs. The key contributions of our algorithm are twofold:

\textbf{We design an assignment that minimizes the variance of the triangle count in Section \ref{sec:mincov}}
Each triangle is incident to three nodes. It is known that assigning each triangle to a carefully chosen node and letting only that node report it can greatly reduce the variance and thus the error in triangle counting \cite{hillebrand_et_al:LIPIcs.STACS.2025.49,vldb2025}. 
In the unweighted setting, however, the server does not know the graph structure, so prior work relied on indirect statistics such as noisy degree estimates to guide this assignment. In contrast, under differential privacy for weighted graphs, the server is allowed to know the graph structure. 
By exploiting this additional information, we construct an optimal max-flow min-cost algorithm designed to minimize counting variance. For scenarios where this optimal approach is computationally impractical, we also provide an efficient and scalable approximation algorithm. 

Furthermore, we prove lower bounds on the precision of any local weight differentially private algorithm operating on planar and degeneracy-bounded graphs. Because these bounds also apply to the local edge differential privacy setting, they establish theoretical lower bounds for the problem studied in \cite{hillebrand_et_al:LIPIcs.STACS.2025.49}. 
Finally, we demonstrate that with the optimal assignment, a variant of our algorithm achieves asymptotically optimal precision on planar graphs and bounded-degeneracy graphs.

\textbf{We develop a scalable algorithm to compute the smooth sensitivity in Section \ref{sec:smooth}.}
Smooth sensitivity \cite{nissimsmooth} is a standard tool for reducing the amount of noise added to released statistics and can yield substantial accuracy improvements in the weighted setting. 
However, computing a sufficiently tight smooth sensitivity bound is often computationally expensive in practice \cite{hillebrand2023communication}. Although there are algorithms for computing smooth sensitivity in the unweighted setting, directly applying those bounds to the weighted setting would drastically overestimate the sensitivity and therefore inject far more noise than necessary.
By exploiting the knowledge of graph structure, we develop an algorithm that computes the local \(\beta\)-smooth sensitivity in \(O(d^2 \log^2 d)\) time for a node of degree \(d\).
The key idea is to maintain a sorted array so that we can capture the relevant local changes without recomputing the sensitivity from scratch.
We remark that in typical road or communication networks node degrees are generally well below \(10^4\).  This implies that our smooth sensitivity computation can be carried out within practical running time and is therefore scalable in these settings.

In addition to these two main contributions, in Section~\ref{sec:algo}, we propose to use the unbiased estimator from \cite{unbiased_histo}, originally proposed for privately releasing histograms, so that it can be used to release an unbiased estimate of the number of triangles under our weighted-graph model. 
When we use the biased estimator, the resulting error scales linearly with the number of triangles in the input graph. In contrast, with the unbiased estimator, the error scales on the order of the square root of the number of triangles. Consequently, for graphs with many triangles, our algorithm achieves a significantly smaller relative error.

In Section \ref{sec:experiments}, we evaluate our method on two real-world graphs: a telecommunication network in Milan, Italy, a biological network used for integrated transcriptional and metabolic analysis~\cite{loboda2016solving, barlacchi2015multi}, synthetic complete and planar graphs, and the Google web graph. Compared to a naive baseline that first releases all edge weights using the Laplace mechanism and then counts triangles from the noisy graph, our algorithm achieves up to two orders of magnitude smaller relative counting error.

\section{Preliminaries}\label{sec:prelim}

\subsection{Problem definition}
Let \(G=(V,E,w)\) be a weighted undirected graph with vertex set \(V=\{1,\dots,n\}\) and edge set \(E\subseteq \binom{V}{2}\).
For each edge \(\{u,v\}\in E\), let \(w_{uv}=w_{vu}\in\mathbb{Z}\) denote its (integer) weight; we restrict to integer weights, which suffice for all applications considered in this paper.
For a vertex \(v\in V\), write \(\mathcal{N}(v)=\{u\in V:\{u,v\}\in E\}\) for its neighborhood and \(d_v=|\mathcal{N}(v)|\) for its degree. The maximum degree is denoted by \(d_{max}\).
The incident–weight vector at \(v\) is \(w^{v}=(w_{vu})_{u\in \mathcal{N}(v)}\in\mathbb{Z}^{d_v}\).
Throughout, the graph structure \((V,E)\) is public, whereas each weight \(w_{uv}\) is known only to its endpoints \(u\) and \(v\); accordingly, \(w^{v}\) constitutes the private data of node \(v\).

With \(\Delta\) we denote the set of triangles of \(G\) where a triangle \(T \in \Delta\) is either represented by the set of three edges or the set of three nodes forming the triangle.
We use the edge and node representation of a triangle interchangeably and clarify the used representations in statements. We now present the formal definition of counting below-threshold triangle counts.

\begin{definition}
    Given a weighted undirected graph \(G = (V,E,w)\) and a threshold \(\lambda \in \mathbb{Z}\), the goal is to output 
        \(f(G) = \sum_{\left\{v, u, x\right\} \in \Delta} 1\{w_{vu} + w_{vx} + w_{ux} < \lambda\}.\)
\end{definition}

\subsection{Local weight differential privacy}
In this section, we introduce the formal notations for our privacy setting together with a discrete and continuous privacy mechanism.
In the setting of local weight differential privacy, it is assumed that the graph topology, nodes and edges, are public and the private information we want to protect is encoded in the edge weights.
This assumption is considered in several previous works such as \cite{sealfon2016,zou2025almost}, and is natural in applications such as road and telecommunication networks, where the underlying connectivity (streets, intersections, or network links) is typically public, while sensitive usage information is encoded in the edge weights.

\begin{definition}[Neighboring weight vectors]
    We define two weight vectors \(w, w' \in \mathbb{Z}^d\) as neighboring \(w \sim w'\) if their \(l_1\) distance is at most one, i.e.
       \(w \sim w' \iff ||w - w'||_1 \le 1\)
\end{definition}

\begin{definition}[Local weight differential privacy]
    Let \(\varepsilon > 0\). A randomized query \(\mathcal{R}\) satisfies \(\varepsilon\)-local weight differential privacy (\(\varepsilon\)-LWDP) if, for any possible neighboring weight vectors \(w\), \(w'\), and any possible outcome set \(S\), 
    \(\Pr[\mathcal{R}(w) \in S] \leq e^{\varepsilon} \Pr[\mathcal{R}(w') \in S]\).

An algorithm \(\mathcal{A}\) is said to be \(\varepsilon\)-LWDP if, for any node \(v_i\), and any sequence of queries \(\mathcal{R}_1, \dots, \mathcal{R}_\kappa\) issued to \(v_i\), where each query \(\mathcal{R}_j\) satisfies \(\varepsilon_j\)-local weight differential privacy (for \(1 \leq j \leq \kappa\)), the total privacy loss is bounded by \(\varepsilon_1 + \dots + \varepsilon_\kappa \leq \varepsilon\). 
We call \(\varepsilon\) the \emph{privacy budget} of any \(\varepsilon\)-LWDP algorithm and query. \end{definition}

By the definition of LWDP, an attacker cannot distinguish between two weight vectors that differ by one unit. 
Since each weight represents the number of mobility or communication events in a road or telecommunication network, this implies that an attacker cannot determine whether any specific mobility or communication event occurred in the database.

We will now introduce two private queries which we will use in our mechanism. The first one is the discrete Laplace query.
\begin{definition}[Discrete Laplace query \cite{ghosh2009universally}]\label{def:dlap}
    For \(p \in (0,1)\), \(DLap(p)\) denotes the discrete Laplace distribution with probability mass at \(i \in \mathbb{Z}\) of \(\frac{1-p}{1+p} \cdot p^{|i|}\).
    For a weight vector \(w = (w_1, \dots, w_d) \in \mathbb{Z}^d\), the discrete Laplace privacy mechanism adds independently drawn noise from \(DLap(e^{-\varepsilon})\) to each entry in \(w\), i.e. 
       \(\tilde{w}_i = w_i + DLap(e^{-\varepsilon}) \quad \forall i \in [d]\).
    The discrete Laplace query also known as geometric query satisfies \(\varepsilon\)-LWDP.
\end{definition}

The second query is the Laplace query, whose definition relies on global sensitivity.

\begin{definition}[Global sensitivity \cite{dwork}]\label{def:gsens}
    For a function \(f: \mathbb{Z}^d \rightarrow \mathbb{R}\), the global sensitivity \(GS(f)\) is defined as     \(GS(f) := \max_{w \sim w'} ||f(w) - f(w')||_1.\)
\end{definition}
\begin{definition}[Laplace query \cite{dwork}]\label{def:rel_lap}
    For a function \(f: \mathbb{R}^d \rightarrow \mathbb{R}\) and \(\varepsilon > 0\), the following is defined as the Laplace query
        \(LM_f(w) = f(w) + (Y_1, \dots, Y_d)\) 
    where \(Y_1, \dots, Y_d\) are drawn from \(Lap(GS(f) / \varepsilon)\). 
The Laplace query satisfies \(\varepsilon\)-local weight differential privacy. 
\end{definition}

\section{Two-step algorithm for publishing below-threshold triangle counts} \label{sec:algo}
In this section, we present our two-step algorithm to release the number of below-threshold triangles under local weight differential privacy. 
Algorithm~\ref{alg:basis} proceeds in two rounds of communication, following the two-step paradigm of~\cite{imola_2step}.
In the first step, nodes publish their incident weight vectors such that the central server can then construct a noisy graph with this information.
In the second step, nodes locally use their private weights together with the noisy graph to accurately estimate the local number of below-threshold triangle counts. 
These local counts are published under local weight differential privacy and then aggregated by the central server.

\begin{algorithm}
\caption{\textsc{Two-step algorithm for counting below-threshold triangles}}
\label{alg:basis}
\begin{algorithmic}[1]
\Statex \hspace{-4mm} \textbf{Input:} weighted graph \(G = (V, E, w)\), \(\lambda\), \(\varepsilon_1\), \(\varepsilon_2\) 
\Statex \hspace{-4mm} \textbf{Output:} below-threshold triangle count \(\tilde{k}\) 

\Statex \hspace{-4mm} \underline{First step}: Noisy releases and task assignments

\Statex \hspace{-4mm} \textbf{Node} \(v \in V\):
\State \(\tilde{w}^v \gets w^v + (Y_1, \dots, Y_{d_v})\) where \(Y_i = DLap(e^{-\varepsilon_1})\) for all \(i \in [d_v]\). \Comment{Discrete Laplace query}
\State Send \(\tilde{w}^v\) to the server
\Statex \hspace{-4mm} \textbf{Server:}
\State Collect \(\tilde{w}^v\) from each \(v \in V\)
\State Construct \(G' = (V, E, w')\) from \((\tilde{w}^v)_{v \in V}\).  \Comment{Section \ref{subsec:step1}}
\State Compute assignment \(\rho: \Delta \mapsto V\) from \((V,E)\).  \Comment{Section~\ref{sec:mincov}}
\State Distribute \(G', \rho\) to each node
\Statex
\Statex \hspace{-4mm} \underline{Second Step}: Local counting
\Statex \hspace{-4mm} \textbf{Node} \(v \in V\):
\State Let \(\Delta_v = \rho^{-1}(v)\).  \(f_v'(w^v) \gets \sum_{T \in \Delta_v} g_v^T(w^v)\) when  \(g_v^T(w^v)\) is the estimator declaring if \(T\) is below the threshold.\Comment{Section \ref{subsec:step2}}
\State Compute the sensitivity of \(f'_v\), denoted \(s_v\).
\Comment{Theorem \ref{thm:sensitivity} or Section \ref{sec:smooth}}
\State Compute \(\tilde{f}_v(w^v)\), which is the value of \(f'_v(w^v)\) released using either the Laplace query (Definition~\ref{def:rel_lap}) or the smooth-sensitivity query (Definition~\ref{def:smooth_query}). We set the privacy budget of this release to~\(\varepsilon_2\).
\State Send \(\tilde{f}_v(w^v)\) to the server
\Statex \hspace{-4mm} \textbf{Server:}
\State \Return  \(\tilde{k} \leftarrow \sum_{v \in V} \tilde{f}_v(w^v)\)
\end{algorithmic}
\end{algorithm}

\subsection{Noisy releases and task assignments}
\label{subsec:step1}
Each vertex \(v\) releases its incident–weight vector using the discrete Laplace query (Definition~\ref{def:dlap}), yielding
\(\tilde{w}^{v}=(\tilde{w}_{vu})_{u\in \mathcal{N}(v)}\) for all \(v \in \{1, \dots, n\}\).
A central server aggregates the reports and constructs a noisy graph \(G'=(V,E,w')\) by enforcing symmetry via an index tie-break: \(w'_{uv}=\tilde{w}_{uv}\) if \(u<v\), and \(w'_{uv}=\tilde{w}_{vu}\) otherwise, for all \(\{u,v\}\in E\).

Let \(\Delta\) denote the set of triangles in \(G\). The server assigns each triangle \(T \in \Delta\)
to one of its vertices via a mapping \(\rho:\Delta \to V\). For each node \(v \in V\), let
\(\Delta_v = \rho^{-1}(v)\) denote the set of triangles for which \(v\) is responsible. In
Section~\ref{sec:mincov}, we define \(\rho\) to minimize the expected error. The server then
notifies each node \(v\) of its assigned set \(\Delta_v\). Moreover, for every triangle
\(T=\{v,u,x\}\in \Delta_v\), the server transmits to \(v\) the noisy weight \(w'_{ux}\) of the
non-incident edge \(\{u,x\}\).

\subsection{Local counting}
\label{subsec:step2}
Each node \(v\) locally counts the number of below-threshold triangles among its assigned set
\(\Delta_v=\rho^{-1}(v)\).
At this stage, \(v\) knows the public topology \((V,E)\), its private incident-weight vector \(w^{v}\), and the noisy
weights \(w'_{u,x}\) for all triangles in \(\{u,v,x\} \in \Delta_v\) released in Step~1.
We consider two estimators for declaring whether \(T\) is below the threshold; in
Algorithm~\ref{alg:basis} this is represented by the placeholder \(g_v^T(w^v)\).
Node \(v\) applies the chosen estimator to each \(T\in\Delta_v\) and counts those declared below threshold.
We first introduce a biased estimator.
\begin{definition}[Biased estimator]
   Let \(v\in V\) be the vertex responsible for a triangle \(T=\{v,u,x\}\in \Delta_v\).
We define the following \emph{biased} estimator:
  \(B_T' \;=\; \mathbf{1}\!\left\{\, w_{vu} + w_{vx} + w'_{ux} < \lambda \,\right\}\),
where \(w_{vu}\) and \(w_{vx}\)  are true edge weights incident to \(v\) and \(w'_{ux}\) is the noisy weight from Step~1.
\end{definition}

We next recall the \emph{unbiased} estimator of \cite{unbiased_histo}, originally proposed for releasing anonymous histograms without bias.
\begin{definition}[Unbiased estimator]
  Let \(h:\mathbb{Z}\to\mathbb{R}\) be defined by setting \(h(m)=0\) if \(m>\lambda\), \(h(m)=-p/(1-p)^2\) if \(m=\lambda\), \(h(m)=1+p/(1-p)^2\) if \(m=\lambda-1\), and \(h(m)=1\) if \(m<\lambda-1\), where \(p=e^{-\varepsilon_1}\).
    Let \(v \in V\) be the vertex responsible for counting the triangle \(T = \{v,u,x\} \in \Delta_v\). Then, with \(U'_T\) we define the following estimator:
    \(U'_T = h(w_{vu} + w_{vx} + w'_{ux}).\)
\end{definition}
Both estimators \(B'_T\) and \(U'_T\) use the true incident weights \(w_{vu}\) and \(w_{vx}\) together with the noisy value \(w'_{ux}\).
Define \(f'_v\) as the local below-threshold count at node \(v\), with \(g\) a placeholder for the chosen estimator: \(f'_v(w^v) \;=\; \sum_{T\in \Delta_v} g_v^T\left(w^v\right).\)

We set \(g^T_v(w^v)=B'_T\) when using the biased estimator; otherwise, we set \(g^T_v(w^v)=U'_T\).

\subsection{Publishing local counts}
Because computing the local below-threshold count \(f'_v(w^v)\) depends on the private incident-weight vector \(w^v\), we must apply a privacy mechanism to ensure local weight differential privacy before sending \(f'_v(w^v)\) to the central server. Note that the vector \(w'_v\) from Step~1 is already public and thus treated as a constant here. Since the final output does not need to be an integer, we use the Laplace query of Definition~\ref{def:rel_lap}. The proof of all theorems in this section can be found in Appendix \ref{appendix:analysis}.

\begin{theorem}[Global sensitivity of \(f_v'\)] \label{thm:sensitivity}
    Let \(\Delta_v\) denote the set of triangles \(v\) is responsible for.
    The global sensitivity of \(f'_v\) is given by
    \(
        GS(f'_v) = GS(g^T_v) \max_{u:\{v, u\} \in E} |\{T \in \Delta_v: \{v, u\} \in T\}|
    \)
    where \(GS(g^T_v) = 1\) when using the biased estimator and \(GS(g^T_v) = 1 + 2\frac{p}{(1-p)^2}\) when using the unbiased estimator. \label{thm:global}
\end{theorem}
The computation of \(GS(f'_v)\) depends only on public data.
Therefore, each node \(v\in V\) can compute the sensitivity \(GS(f'_v)\) locally. 
For \(v\in V\) with degree \(d\), the quantity \(\bigl|\{T\in\Delta_v:\, e\in T\}\bigr|\) can be computed in \(O(d)\) time by scanning each other edge incident to \(v\) and checking whether it closes a triangle. Repeating this for all incident edges yields a total running time of \(O(d^2)\).

\subsection{Analysis of two-step algorithm}\label{sec:analysis}

In the following we proof local weight differential privacy and analyze the communication costs and precision of Algorithm~\ref{alg:basis}.

\begin{theorem} [Privacy of Algorithm \ref{alg:basis}]
    Algorithm \ref{alg:basis} satisfies \((\varepsilon_1 + \varepsilon_2)\)-local weight differential privacy.
\end{theorem}
\begin{proof}
    Algorithm~\ref{alg:basis} issues two queries to each node.
The first retrieves the incident weight vector of a node \(v\in V\); \(v\) responds using the discrete Laplace query, achieving \(\varepsilon_1\)-LWDP.
The second retrieves the local below-threshold count; here the Laplace query is used, achieving \(\varepsilon_2\)-LWDP.
\end{proof}

\begin{theorem} [Communication cost of Algorithm \ref{alg:basis}]\label{thm:comm_costs}
    With probability at least \(1 - \frac{1}{m}\), the communication cost of Algorithm \ref{alg:basis} is \(\tilde{O}(m + |\Delta|)\).
\end{theorem}

We now bound the expected squared \(\ell_2\)-error of the unbiased and biased estimator. The error arises from the bias and variance of the estimator \(g^T_v\) and the Laplace query for publishing the local counts. The global sensitivity from Theorem~\ref{thm:global} is bounded by the degree of a vertex resulting in a dependency on \(d_{max}\) in the error bound. 
For two vertices \(x, y\in V\), let \(N_{x,y} = \{v \in V : \{v, x, y\} \in \Delta_v\}\) denote the set of vertices that form a triangle with \(x, y\) and use the noisy information \(w'_{xy}\). 
For \(v, u \in N_{x,y}\) both use the same noisy information \(w'_{xy}\) which results in a covariance between the estimator with \(T = \{v,x,y\}\) and with \(T'= \{u,x,y\}\). We call this structure a \(C'_4\) instance. Each instance of \(C'_4\), represented by the pair of triangles \(T\) and \(T'\), induces covariance between the counts at nodes \(u\) and \(v\). This covariance is one of the dominant terms in the counting error. Consequently, the error scales with the number of such instances, denoted by \(\#C'_4\).

\begin{theorem}[Precision of biased estimator]\label{thm:bias_err}
    The expected squared \(\ell_2\)-error of Algorithm \ref{alg:basis} using the biased estimator is bounded by 
        \(O\left(\frac{d_{max}m}{\varepsilon^2_2} + \frac{|\Delta|^2}{\varepsilon^2_1} + \frac{\#C'_4}{\varepsilon_1}\right).\)\label{thm:final-biased}
\end{theorem}

\begin{theorem} [Precision of unbiased estimator] \label{thm:final-unbiased}
    The expected squared \(\ell_2\)-error of Algorithm \ref{alg:basis} with the unbiased estimator is bounded by
        \(O\left(\frac{d_{max}m}{\varepsilon^4_1\varepsilon^2_2} + \frac{|\Delta|}{\varepsilon^3_1} + \frac{\#C'_4}{\varepsilon^3_1}\right)\).
\end{theorem}
We observe that the unbiased estimator achieves a smaller error when the number of triangles \( |\Delta| \) is large, whereas the biased estimator is more accurate when the privacy budgets \(\varepsilon_1\) and \(\varepsilon_2\) are small.

\section{Assignment function \(\rho\)}\label{sec:mincov}
In this section, we provide an algorithm that computes an assignment from triangles to nodes \(\rho: \Delta \mapsto V\) to minimize the number of \(C'_4\) instances, which is denoted by \(\#C'_4\).
Let \(l(e)\) define the \emph{load} of an edge \(e = \{u,v\} \in E\) representing the number of estimators that use the noisy weight \(w'_{uv}\), i.e. the number of triangles \(T \in \Delta\) with \(u,v \in T\) such that \(\rho(T) \notin \{u,v\}\).
We define the assignment costs \(c(\rho)\), as the number of \(C'_4\) instances generated by \(\rho\): \(c(\rho) = \#C'_4 = \sum_{e\in E} \binom{l(e)}{2}.\)
We need to assign every triangle \(T\) to one of its edges such that \(\#C'_4\) is minimized. Since the sum of loads \(\sum_{e \in E}l(e)\) is always equal to the number of triangles \(|\Delta|\) minimizing \(c(\rho)\) is equivalent to minimizing the sum of squared loads.
This is formalized in the following optimization problem. 
For a triangle \(T \in \Delta\) and edge \(e \in T\), let \(x_{T,e} \in \{0,1\}\) denote whether \(T\) is assigned to \(e\). We calculate \((x_{T,e})_{T \in \Delta, e \in T} \in \{0,1\}^{|\Delta|\times 3}\) such that \(\sum_{e \in T} x_{T,e} = 1\) for all \(T \in \Delta\) and \(\sum_{e\in E}\left(\sum_{T \in \Delta : e\in T} x_{T,e}\right)^2\) is minimized. 

\begin{theorem}\label{thm:assgn_time}
    Given a graph \(G = (V,E)\), the optimal assignment can be computed in \(O(|\Delta|^2 + |\Delta|(|\Delta| + m) \log(|\Delta| + m))\) time.
\end{theorem}

All proofs for this section are deferred to Appendix~\ref{appendix:assignment}. In dense graphs, the number of triangles \(|\Delta|\) scales with \(O(n^3)\) and computing the optimal assignment becomes impractical. 
We therefore introduce a greedy approach that iterates over all triangles and assigns a triangle to the edge with the currently lowest load.
\begin{theorem} [Approximation ratio of greedy algorithm]\label{thm:greedy_approx}
    Let \(\#C'_{4, ALG}\) be the number of \(C'_4\) instances produced by the greedy algorithm, and \(\#C'_{4, OPT}\) be the number of instances in the optimal assignment. The greedy algorithm achieves the following bound
    \(\#C'_{4, ALG} \le (3 + 2\sqrt{2})\#C'_{4, OPT} + (1 + \sqrt{2})|\Delta|.\)
\end{theorem}
We can now strengthen Theorems \ref{thm:final-biased} and \ref{thm:final-unbiased} replacing \(\#C_4'\) with \(\#C'_{4,OPT}\). This refinement is possible because we operate under local weight differential privacy. In earlier work on local differential privacy, such as \cite{hillebrand_et_al:LIPIcs.STACS.2025.49}, such an improvement is not achievable, since the server does not know the graph structure.

\textbf{Assignment function for planar graphs:} We now discuss the assignment when the input graph is planar.
\begin{lemma}\label{lem:planar_opt}
Let \(G\) be a planar graph. Then the optimal assignment satisfies \(\#C'_{4,{\mathrm{OPT}}} = 0\). Moreover, such an optimal assignment can be found in \(O(n^2\log n)\) time.
\end{lemma}

\begin{lemma}\label{lem:planar_lower}
    There exists a planar graph \(G\) with \(n\) nodes such that the expected squared \(\ell_2\)-error of any \(\varepsilon\)-LWDP algorithm is in \(\Omega (n^2)\).
\end{lemma}

\begin{theorem}
Algorithm~\ref{alg:basis}, instantiated with the biased estimator and the assignment function from Lemma~\ref{lem:planar_opt}, achieves asymptotically optimal precision on planar graphs.
\end{theorem}
\begin{proof}
    For planar graphs, \(\Delta = O(n)\), \(d_{max} \le n\), and \(\#C'_4 = 0\). Plugging this into Theorem~\ref{thm:bias_err} with \(\varepsilon_1 = \varepsilon_2 = \varepsilon/2\) yields an upper bound of \(O(n^2/\varepsilon^2)\) while Lemma~\ref{lem:planar_lower} provides the lower bound.
\end{proof}

\textbf{Assignment function for bounded-degeneracy  graphs:} In the following, let \(\gamma\) denote the degeneracy of the input graph. Triangle counting under local edge differential privacy (LEDP) on bounded-degeneracy graphs was analyzed in~\cite{hillebrand_et_al:LIPIcs.STACS.2025.49}. However, because the structures inducing covariance differ slightly between LEDP and LWDP, the results of~\cite{hillebrand_et_al:LIPIcs.STACS.2025.49} do not directly carry over. We therefore provide proofs tailored to the LWDP setting.
In addition, we establish two separate lower bounds under LWDP and LEDP, respectively, addressing whether the method of~\cite{hillebrand_et_al:LIPIcs.STACS.2025.49} can be improved.
\begin{lemma}\label{lem:deg_ass}
    Let \(G\) be a graph with degeneracy \(\gamma\), then there exists a triangle-node assignment with \(c(\rho) = O(n\gamma^3)\).
\end{lemma}

\begin{lemma}\label{lem:deg_err}
    For a graph \(G\) with degeneracy \(\gamma\), the expected squared \(\ell_2\)-error of Algorithm~\ref{alg:basis} with unbiased estimator and privacy budget \(\varepsilon_1, \varepsilon_2\) is bounded by
        \(O(\frac{n^2\gamma}{\varepsilon^4_1\varepsilon^2_2} + \frac{n\gamma^3}{\varepsilon^3_1})\).
\end{lemma}
\begin{proof}
    \(|\Delta|\) is upper bounded by \(O(n\gamma^2)\), \(m = O(n\gamma )\), \(d_{max} \le n\), and by Lemma~\ref{lem:deg_ass} there exists an assignment such that \(\#C'_{4} \in O(n\gamma^3)\). The claim then follows from Theorem~\ref{thm:bias_err}.
\end{proof}

We now establish lower bounds for the expected squared \(\ell_2\)-error on graphs with degeneracy \(\gamma\). Our analysis adapts the construction from \cite{eden2025triangle}, which proved an \(\Omega(n^3/\varepsilon^2)\) lower bound for triangle counting under local edge differential privacy, via a reduction from computing the sum of either \(\gamma\) or \(n-\gamma\) values.

\begin{theorem}\label{thm:deg_lb}
    There exist families of graphs with degeneracy \(\gamma\) such that any \(\varepsilon\)-LWDP algorithm outputting the number of below-threshold triangles with an additive error of \(\alpha\) with probability at least \(2/3\) must satisfy
    \(\alpha \in \Omega\left( \frac{1}{\varepsilon} \max\bigl\{ (n-\gamma)\sqrt{\gamma}, \, \gamma\sqrt{n-\gamma} \bigr\} \right).\)
\end{theorem}

Theorems \ref{thm:deg_lb} shows that the error achieved by the unbiased estimator is close to optimal. The \(O(n^2\gamma)\) dependency is tight according to Theorem~\ref{thm:deg_lb} while there remains a \(\gamma\) factor gap between the \(O(n\gamma^3)\) term and the bound from and the \(\Omega(n\gamma^2)\) of Theorem~\ref{thm:deg_lb}. In Appendix \ref{subsec:exp_assign} we provide empirical results on the running times and costs of the assignment algorithms proposed in this section.

\section{Efficient smooth sensitivity calculation}\label{sec:smooth}
In this section, we will use smooth sensitivity to ensure privacy during the second step of our algorithm. 
The second step publishes the local below-threshold count of \(v \in V\). Previously, we saw the global sensitivity is in \(O(d_v)\).
This is due to the fact that the increase or decrease of a single edge weight affects at most \(d_v - 1\) triangles and, therefore, might change all triangle weights from being less than \(\lambda\) to being at least \(\lambda\) or vice versa in the worst case. 
However, in most cases the sensitivity is way smaller. 
To capture this we use the following notion of local sensitivity.
\begin{definition}[Local sensitivity \cite{nissimsmooth}]
    For a function \(f: \mathbb{Z}^d \rightarrow \mathbb{R}\) and a weight vector \(w \in \mathbb{Z}^d\), the local sensitivity of \(f\) at \(w\) is
    \(
        LS_f(w) = \max_{y:y \sim w} |f(w) - f(y)|
    \)
\end{definition}

Local sensitivity depends on the input weight vector \(w^v \in \mathbb{Z}^{d_v}\) associated with a vertex \(v \in V\) of degree \(d_v\). 
The \(\beta\)-smooth sensitivity is defined as follows:

\begin{definition}[\(\beta\)-Smooth Sensitivity \cite{nissimsmooth}]
    For \(\beta > 0\), the \(\beta\)-smooth sensitivity of \(f\) at \(w\) is defined as 
    \(
        S^*_{f,\beta}(w) = \max_{y \in \mathbb{Z}^d}\left(LS_f(y)e^{-\beta \cdot d(w,y)}\right)
    \)
\end{definition}

We then give the definition of the smooth sensitivity query below.
\begin{definition}[\(\beta\)-smooth sensitivity query \cite{nissimsmooth, smooth_sensitivity}]\label{def:smooth_query}
    For a function \(f\), \(\varepsilon_2 > 0\) and \(\Gamma > 1\), the following is defined as the \(\beta\)-smooth sensitivity query
        \(SM_f(w) = f(w) + \frac{2(\Gamma -1)^{\frac{\Gamma - 1}{\Gamma}}}{\varepsilon_2}S^*_{f, \varepsilon_2/2(\Gamma -1)}(w) \cdot Z\)
    where \(Z\) is drawn from a probability distribution proportional to \(1/(1+|z|^\Gamma)\).
    The smooth sensitivity mechanism satisfies \(\varepsilon_2\)-differential privacy.
\end{definition}

The mechanism remains valid for every \(\Gamma > 1\). However, for \(\Gamma \le 3\) we currently do not know how to express \(\operatorname{Var}[Z]\) in closed from~\cite{hillebrand2023communication}. In this paper, we fix \(\Gamma = 4\), a choice for which \(\operatorname{Var}[Z] = 1\).

We now present our algorithm to compute the smooth sensitivity of the local count \(f'_v\) for both the biased and unbiased estimators. 
The core idea is to fix the edge whose weight increases or decreases during the local sensitivity computation, and then allocate weights such that this local sensitivity is maximized while balancing the penalty term. 
We demonstrate that by maintaining a sorted array of partial triangle weights, this weight distribution step, a bottleneck operation that has to be conducted \(d\) times, can be carried out in \(O(d \log^2 d)\) time. 
This is a significant improvement over the straightforward \(O(d^2)\) approach. 
Full details are deferred to Appendices~\ref{appendix:smooth_bias} and~\ref{appendix:smooth_unbiased}.

\begin{theorem}[Computation time of smooth sensitivity]\label{thm:smooth_run_time}
    There exists an algorithm that computes the \(\beta\)-smooth sensitivity of \(f'_v\) for a node \(v\) with degree \(d\) in \(O(d^2 \log^2 d)\) time.
\end{theorem}

This improvement allows us to replace the Laplace query, used in Step 2 to publish local triangle counts, with the \(\beta\)-smooth sensitivity query from Definition~\ref{def:smooth_query}. As demonstrated in Section~\ref{sec:experiments}, this approach yields significant utility benefits while our proposed algorithm ensures that the overall computation time remains practical.

\section{Experiments}\label{sec:experiments}
In this section, we evaluate four variants of Algorithm~\ref{alg:basis} on a graph (ML-Tele-278) derived from telecommunication data collected in the city of Milan, Italy. The details of the data to graph conversion are given in Appendix \ref{appendix:experiments}.
The implementation used in our experiments is publicly available\footnote{Source code: \url{https://github.com/Crightub/private-below-threshold-triangle-counting}}. 
All experiments were run on the SHIROKANE supercomputer, provided by the Human Genome Center, Institute of Medical Science, The University of Tokyo\footnote{For hardware specifications, see \url{https://supcom.hgc.jp/english/sys_const/system-main.html}.}.

We evaluate four variants of Algorithm~\ref{alg:basis}: \emph{global-biased}, \emph{global-unbiased}, \emph{smooth-biased}, and \emph{smooth-unbiased}.
The \emph{biased} variants apply Algorithm~\ref{alg:basis} with the biased estimator \(B'_T\) to compute the local below-threshold triangle counts, while the \emph{unbiased} variants use the unbiased estimator \(U'_T\).
For Step~2, the \emph{global} variants apply the Laplace query from Definition~\ref{def:rel_lap}, and the \emph{smooth} variants apply the smooth-sensitivity query from Definition~\ref{def:smooth_query}. In all cases, we use the approximation algorithm (Theorem \ref{thm:greedy_approx}) to compute the assignment function \(\rho : V \mapsto \Delta\).
We also compare these methods against a non-interactive baseline, referred to as \emph{baseline}. In this approach, the central server requests the incident weight vector of each node \(v\) using the discrete Laplace mechanism, and then estimates the below-threshold triangle counts solely from the resulting noisy weights. We fix \(\lambda = 4\) and we use a total privacy budget of \(\varepsilon = 2\) for both the baseline and our method; in our method we split this as \(\varepsilon_1 = \varepsilon_2 = 1\). A total privacy budget of \(2\) is standard in prior work on graph differential privacy~\cite{imola_2step,hillebrand2023communication}.
In our experiments, we executed each algorithmic variant 10 times and report the average relative error. The relative error is defined as \(|r - p| / \mathsf{r}\), where \(r\) is the true number of triangles below the threshold and \(p\) is the number of triangles reported by the algorithms. We selected 10 repetitions because this was sufficient to obtain stable estimates across all experimental settings we evaluated.

\begin{figure*}
    \centering
    \begin{tikzpicture}

    \begin{groupplot}[
        group style={
            group size=3 by 1,
            horizontal sep=1.2cm,
            vertical sep=1cm
        },
        width=0.3\textwidth,
        height=0.23\textwidth,
        grid=both,
        grid style={dotted,gray!30},
        tick label style={font=\footnotesize},
        label style={font=\footnotesize},
        legend style={
            font=\footnotesize,
            /tikz/every even column/.style={column sep=6pt},
            draw=none,
            fill=none,
            at={(0.5,1.15)},
            anchor=south,
            legend columns=3
        },
        legend cell align={left},
        yminorticks=true,
    ]

    \nextgroupplot[
        xlabel={\(\varepsilon\)},
        ylabel={relative error},
        ymode=log,
        ytick={1e1,1e-0,1e-1,1e-2,1e-3,1e-4},
        yticklabels={\(10^{1}\),\(10^{0}\),\(10^{-1}\),\(10^{-2}\),\(10^{-3}\),\(10^{-4}\)},
        minor y tick num=8, 
    ]

\node[anchor=north west, font=\footnotesize] at (rel axis cs:0.5,1.00) {(a)};

    \addplot+[thick, mark=*, mark options={solid}, color=blue] table [x=x, y=smooth_unbiased_l2_rel, col sep=comma] {fig/eps_rel_error.csv};
    \addlegendentry{smooth-unbiased}

    \addplot+[thick, mark=square*, mark options={solid}, color=red] table [x=x, y=naive_l2_rel, col sep=comma] {fig/eps_rel_error.csv};
    \addlegendentry{baseline}

    \addplot+[thick, mark=o, color=orange] table [x=x, y=global_unbiased_l2_rel, col sep=comma] {fig/eps_rel_error.csv};
    \addlegendentry{global-unbiased}

    \addplot+[thick, mark=square, color=green] table [x=x, y=global_biased_l2_rel, col sep=comma] {fig/eps_rel_error.csv};
    \addlegendentry{global-biased}

    \addplot+[thick, mark=diamond*, mark options={solid}, color=purple] table [x=x, y=smooth_biased_l2_rel, col sep=comma] {fig/eps_rel_error.csv};
    \addlegendentry{smooth-biased}

    \nextgroupplot[
        xlabel={\(|\Delta|\)},
        ylabel={},
        yticklabel pos=right,
        ymode=log,
        ymin=1e-3,
        ymax=1e-1,
         ytick={1e-1,1e-2,1e-3},
        yticklabels={\(10^{-1}\),\(10^{-2}\),\(10^{-3}\)},
        minor y tick num=1,
        ytick distance=200, 
    ]
\node[anchor=north west, font=\footnotesize] at (rel axis cs:0.5,1.00) {(b)};

    \addplot+[thick, mark=*, mark options={solid}, color=blue] table [x=x, y=smooth_unbiased_l2_rel, col sep=comma] {fig/size_rel_error.csv};

    \addplot+[thick, mark=square*, mark options={solid}, color=red] table [x=x, y=naive_l2_rel, col sep=comma] {fig/size_rel_error.csv};

    \addplot+[thick, mark=o, color=orange] table [x=x, y=global_unbiased_l2_rel, col sep=comma] {fig/size_rel_error.csv};

    \addplot+[thick, mark=square, color=green] table [x=x, y=global_biased_l2_rel, col sep=comma] {fig/size_rel_error.csv};

    \addplot+[thick, mark=diamond*, mark options={solid}, color=purple] table [x=x, y=smooth_biased_l2_rel, col sep=comma] {fig/size_rel_error.csv};

    \nextgroupplot[
        xlabel={\(\lambda\)},
        ymode=log,
        ylabel={},
        yticklabel pos=right,
        ytick={1e1,1e-0,1e-1,1e-2,1e-3,1e-4},
        yticklabels={\(10^{1}\),\(10^{0}\),\(10^{-1}\),\(10^{-2}\),\(10^{-3}\),\(10^{-4}\)},
        minor y tick num=8,
    ]
\node[anchor=north west, font=\footnotesize] at (rel axis cs:0.5,1.00) {(c)};

    \addplot+[thick, mark=*, mark options={solid}, color=blue] table [x=x, y=smooth_unbiased_l2_rel, col sep=comma] {fig/lambda_rel_error.csv};

    \addplot+[thick, mark=square*, mark options={solid}, color=red] table [x=x, y=naive_l2_rel, col sep=comma] {fig/lambda_rel_error.csv};

    \addplot+[thick, mark=o, color=orange] table [x=x, y=global_unbiased_l2_rel, col sep=comma] {fig/lambda_rel_error.csv};

    \addplot+[thick, mark=square, color=green] table [x=x, y=global_biased_l2_rel, col sep=comma] {fig/lambda_rel_error.csv};

    \addplot+[thick, mark=diamond*, mark options={solid}, color=purple] table [x=x, y=smooth_biased_l2_rel, col sep=comma] {fig/lambda_rel_error.csv};

    \end{groupplot}

    \end{tikzpicture}

    \caption{Relative error across all methods under different settings for the input graph ML-Tele-278:
    (a) as a function of the privacy parameter \(\varepsilon\),
    (b) as a function of the number triangles \(|\Delta|\),
    (c) as a function of the threshold \(\lambda\).}
    \label{fig:tele_all}
\end{figure*}

Figure~\ref{fig:tele_all} shows the results on \textsc{ML-Tele-278}. The relative error is below \(0.1\) across almost all experiments for all algorithms. Despite the fact that the graph has very high degree and all computations are performed centrally on the server rather than being distributed across nodes, the total runtime remains below five minutes even for the smooth-sensitivity mechanism. This demonstrates that the smooth-sensitivity approach introduced in Section~\ref{sec:smooth} is both scalable and practical for large graphs.

As shown in Figure~\ref{fig:tele_all}(a), our algorithm with the unbiased estimator outperforms all other methods for almost all values of \(\varepsilon\). Consistent with the theoretical analysis, the smooth-sensitivity variant achieves about one order of magnitude smaller relative error than the global-sensitivity variant. The unbiased estimator is only worse than the biased estimator when \(\varepsilon \ll 1\), which matches the theoretical dependence.
We observe that, for \(\varepsilon = 4\), our unbiased algorithm and the baseline achieve comparable relative error. This is because a privacy budget of this size allows all methods to produce highly accurate estimates.

Figure~\ref{fig:tele_all}(b) reports the results as we vary the number of triangles \(|\Delta|\) in the input graph. We uniformly sample vertex subsets of various sizes and run each algorithm on the induced subgraph. Under this sampling procedure, the true triangle count grows linearly in \(|\Delta|\).
The results are consistent with Theorem~\ref{thm:final-biased}, which states that the biased estimator’s squared \(\ell_2\)-error grows quadratically in \(|\Delta|\), resulting in a constant relative \(\ell_2\)-error as depicted.
Theorem~\ref{thm:final-unbiased} shows that the squared \(\ell_2\)-error of the unbiased estimator grows only linearly in \(|\Delta|\). As a result, the relative error of the unbiased estimator is decreasing for growing \(|\Delta|\).

We now evaluate performance as we vary the threshold \(\lambda\), as summarized in Figure~\ref{fig:tele_all}(c). Consistent with the trends in Figures~\ref{fig:tele_all}(a) and \ref{fig:tele_all}(b), we observe roughly an order-of-magnitude improvement from the unbiased estimator, and an additional substantial improvement when using smooth sensitivity.
For large \(\lambda\), the baseline performs nearly as well as our unbiased estimator with smooth sensitivity. This occurs because most triangle weights are very small; even after adding discrete Laplace noise to the edge weights, only a few triangles exceed a large threshold. In such regimes, the two-step algorithms still incur error from the second step, where Laplace noise is added to the below-threshold counts. This introduces a roughly constant error term, which becomes dominant when the true count above the threshold is very small.

In addition to the ML-Tele-278 graph, we have also conducted experiments on a sparse graph produced by GAM, a web service for integrated transcriptional metabolic network analysis \cite{loboda2016solving, sergushichev2016gam}. 
In this setting the baseline algorithm outperforms the two-step approach highlighting the limitations of a two-step approach. The results are shown in Appendix \ref{appendix:experiments}.

\section{Conclusion}
In this work, we investigate the problem of counting below-threshold triangles in weighted graphs under local differential privacy, and we introduce a two-step algorithm to address this challenge.
We develop both biased and unbiased estimators: the biased estimator incurs error that scales linearly with the number of triangles, whereas the unbiased estimator substantially improves accuracy at the expense of a stronger dependence on the privacy budget.
To further improve performance, we incorporate a preprocessing procedure that balances the triangle-counting workload across nodes and can be computed in time linear in the number of triangles.
In addition, we present an efficient algorithm for computing smooth sensitivity, running in \(O(d^{2}\log^{2} d)\) time for a node of degree \(d\), enabling more practical noise calibration.
Our experiments on real-world networks show that the proposed approach consistently outperforms a straightforward baseline, with the largest gains on larger graphs.
These results provide a foundation for accurate subgraph counting in weighted graphs under local differential privacy and suggest natural extensions to more general subgraph statistics.

\section*{Acknowledgment}
Kevin Pfisterer and Quentin Hillebrand are supported by JST SPRING, Grant Number JPMJSP2108. Vorapong Suppakitpaisarn is supported by JST NEXUS Grant Number Y2024L0906031 and
KAKENHI Grant JP25K00369. 
Quentin Hillebrand is part of BARC, Basic Algorithms Research Copenhagen, supported by the VILLUM Foundation grant 54451. Quentin Hillebrand was also supported by a Data Science Distinguished Investigator grant from Novo Nordisk Fonden.

\bibliographystyle{plain}
\bibliography{sources}


\appendix

\section{Analysis of two-step algorithm} \label{appendix:analysis}
In the following we provide the missing proofs for statements made regarding the analysis of Algorithm~\ref{alg:basis} of Section~\ref{sec:algo}.

\begin{proof}[Proof for Theorem~\ref{thm:global}]
    By Definition \ref{def:gsens}, we need to find \(w \sim w'\) such that \(|f'_v(w) - f'_v(w')|\) is maximized. Since \(w\) and \(w'\) are neighboring, there exists exactly one incident edge to \(v\) whose weight differs by \(1\) between \(w\) and \(w'\). 
    Let \(\{v,u\}\) denote this edge. 
    Then, a change of \(w_{vu}\) affects all triangles that contain \(\{v,u\}\) and \(v\) is responsible for counting: \(S = \{T \in \Delta_v: \{v, u\} \in T\}.\)
    In the worst case for each triangle \(T\) in \(S\), the estimator changes by \(GS(g_v^T)\) where \(GS(g_v^T)\) is the maximum change of the respective estimator. It is straightforward to show \(GS(g_v^T)\) for the biased estimator and unbiased estimator. 
\end{proof}

\begin{proof}[Proof of Theorem~\ref{thm:comm_costs}]
The communication incurred by each node consists of three components: (i) uploading the noisy incident-weight vector, (ii) downloading the noisy weights of its assigned triangles, and (iii) uploading its local triangle-count statistics. 

Let \(W\) denote the maximum edge weight in the graph. Using tail bounds for the discrete and continuous Laplace distributions together with a union bound over all released noise variables, we obtain the following results. 

The total communication cost for uploading the noisy weight vectors is bounded by \[ O\left(m \log W + m \log\left(\frac{1}{\varepsilon_1}\log\left(\frac{m}{\gamma}\right)\right)\right) \] with probability at least \(1-\gamma\). During redistribution, only weights corresponding to assigned triangles are transmitted. Since each triangle is assigned to a single node, at most one noisy weight is sent per triangle. The total cost for downloading noisy triangle weights is therefore bounded by \[ O\left(|\Delta| \log W + |\Delta| \log\left(\frac{1}{\varepsilon_1}\log\left(\frac{m}{\gamma}\right)\right)\right) \] with probability at least \(1-\gamma\). Finally, publishing the noisy local triangle counts incurs a communication cost of \[ O\left(|\Delta| \left(\log\left(\frac{1}{\varepsilon_2}\log\frac{|\Delta|}{\gamma}\right) + 1\right)\right) \] with probability at least \(1-\gamma\). We obtain the lemma statement by setting \(\gamma = 1/(3m)\).
\end{proof}

\subsection{Precision of the biased estimator}\label{sec:bias_ana}
We now provide the analysis of the expected squared \(\ell_2\)-error of Algorithm~\ref{alg:basis} under the biased estimator \(B'_T\) proving Theorem~\ref{thm:bias_err}.
Randomness arises in two places: (i) Step~1, where incident weight vectors are released via the discrete Laplace query, and (ii) Step~2, where local below-threshold counts are released via the Laplace query.

\begin{lemma}\label{lem:bias_exp}
    For a triangle \(T \in \Delta\) with triangle weight \(w_T\) the expected value of the estimator \(B'_T\) is given by 
    \[
         \mathbb{E}[B'_T] = \begin{cases}
             1- \frac{p^{\lambda- w_T}}{1+p} &\text{if } w_T < \lambda \\
             \frac{p^{w_T - \lambda + 1}}{1+p} &\text{if } w_T \ge \lambda
         \end{cases}
    \]
\end{lemma}
\begin{proof}
    Let \(T \in \Delta\) with triangle weight \(w_T\). Denote \(w_{vu} + w_{vx} + w'_{ux}\) by \(w'_T\). Notice that \(w'_T = w_{vu} + w_{vx} + w_{ux} + Z = w_T + Z\) where \(Z \sim DLap(p)\). When \(w_T < \lambda\), 
    \[
    \begin{split}
        \mathbb{E}[B_T'] = \Pr[Z < \lambda - w_T] &= \frac{1-p}{1+p}\sum_{i = -\infty}^{\lambda - w_T - 1} p^{|i|} \\
        &= \frac{1-p}{1+p}\left(\sum_{i = 0}^{\infty} p^i - 1 + \sum_{i=0}^{\lambda - w_T -1} p^i\right).
    \end{split}
    \]
    Since \(p = e^{-\varepsilon_1}\), it holds that \(|p| < 1\) and we can apply the closed form solution for the geometric series.
    \[
        \Pr[Z < \lambda - w_T] = 1- \frac{p^{\lambda - w_T}}{1 +p}
    \]
    When \(w_T \geq \lambda\), we apply the same techniques to obtain that
    \[
        \mathbb{E}[B_T'] = \Pr[Z < \lambda - w_T] = \frac{p^{w_T - \lambda +1}}{1+p}.
    \]
\end{proof}

We can imply from Lemma \ref{lem:bias_exp} that the expected value for the triangle \(T \in \Delta\) is independent from the node that is responsible for counting it. 

\begin{lemma} \label{lem:bias}
    The bias for a triangle \(T \in \Delta\) is given by 
    \[
        \mathbb{E}[B'_T] - 1\{w_T < \lambda\} = \begin{cases}
             -\frac{p^{\lambda- w_t}}{1+p} &\text{if } w_T < \lambda \\
             \frac{p^{w_T - \lambda + 1}}{1+p} &\text{if } w_T \ge \lambda

         \end{cases}
    \]
\end{lemma}
\begin{proof}
    The statement directly follows from Lemma \ref{lem:bias_exp}.
\end{proof}

We derive the following two corollaries from Lemma \ref{lem:bias}.

\begin{corollary}\label{cor:bias_bound}
    The bias for a triangle \(T \in \Delta\) is maximized for triangle weights at the boundary cases: \(w_T = \lambda\), \(w_T = \lambda - 1\);
    This results in the following bounds for the bias of the estimator.
    \[
        -\frac{e^{-\varepsilon_1}}{1 + e^{-\varepsilon_1}} \le \mathbb{E}[B'_T] - 1\{w_T < \lambda\} \le \frac{e^{-\varepsilon_1}}{1 + e^{-\varepsilon_1}}.
    \]
\end{corollary}
\begin{corollary}
    For the triangle \(T \in \Delta\), the absolute bias decreases exponentially in \(|\lambda -w_T|\):
    \[
        |\mathbb{E}[B'_T] - 1\{w_T < \lambda\}| \le e^{-\varepsilon_1|\lambda - w_T|}.
    \]
\end{corollary}

Recall that the output of Algorithm \ref{alg:basis} is \(\tilde{k}\) and let \(f(G)\) be the actual count of below-threshold triangles. 
We decompose the expected squared \(\ell_2\)-error into the squared bias plus the variance:
\[
    (\mathbb{E}[\tilde{k}] - f(G))^2 + \mathrm{Var}[f'(G, \varepsilon_1, \varepsilon_2)]
\]

\begin{lemma}\label{lem:bias_squared}
    The squared bias of Algorithm \ref{alg:basis} with biased estimator \(B'_T\) is bounded by 
    \[
        |\Delta|^2\left(\frac{e^{-\varepsilon_1}}{1 + e^{-\varepsilon_1}}\right)^2
    \]
\end{lemma}
\begin{proof}
The expectation is taken over the randomness used when releasing the weight vectors and when releasing the local below-threshold counts.
    \[
        \mathbb{E}[\tilde{k}] = \sum_{v \in V} \mathbb{E}[f'_v(w^v) + Lap(GS(f'_v) / \varepsilon_ 2)]
    \]
    The Laplace query is unbiased. Hence, the bias of the algorithm arises solely from \(f'_v\). 
    \[
        \mathbb{E}[\tilde{k}] - f(G) = \sum_{T \in \Delta} \left( \mathbb{E}[B'_T] - 1\{w_T < \lambda\} \right)
    \]
    From Corollary \ref{cor:bias_bound} the bound on the squared bias follows.
\end{proof}

We observe from the previous lemma that the bias scales the true value by a factor of
\(\frac{e^{-\varepsilon_1}}{1 + e^{-\varepsilon_1}} < 1\).
In other words, the relative error introduced by this biased estimator is this constant factor.

We now analyze the variance of \(\tilde{k}\).
Figure~\ref{fig:cov_ins} visualizes such a \(C'_4\) instance as introduced in Section~\ref{sec:analysis}. 
We use the \(C'_4\) definition to separate the total variance \(\mathrm{Var}[\sum_{v \in V}f_v'(w^v)]\) as follows:
\[
     \sum_{v \in V} \mathrm{Var}[f_v'(w^v)] + \sum_{x,y \in V} \sum_{v,u\in N_{x,y}} \mathrm{Cov}[f_v'(w^v), f'_{u}(w^{u})].
\]

We first analyze the variance of the estimated local below-threshold count.
\begin{lemma}\label{lem:bias_var}
    For a node \(v \in V\), the variance \(\mathrm{Var}[f'_v(G)]\) is bounded by 
    \[
        \frac{e^{-\varepsilon_1}}{1+ e^{-\varepsilon_1}}\left(1- \frac{e^{-\varepsilon_1}}{1+ e^{-\varepsilon_1}}\right)|\Delta_v| = O\left(\frac{|\Delta_v|}{\varepsilon_1}\right)
    \]
\end{lemma}
\begin{proof}
    For every triangle \(T \in \Delta_v\) the noise used in the estimator \(B'_T\) is independent from the noise in the other triangles in \(\Delta_v\). We obtain that:
    \[
        \mathrm{Var}[f'_v(w^v)] = \sum_{T \in \Delta_v} \mathrm{Var}[B'_T]
    \]
    From the fact that \(B'_T\) is an indicator variable and from the probabilities shown in Lemma \ref{lem:bias_exp}, 
    for \(w_T < \lambda\), we obtain that
    \[
        \mathrm{Var}[B'_T] = \left(1- \frac{p^{\lambda - w_T}}{1 + p}\right)\frac{p^{\lambda - w_T}}{1+p}.
    \]
    For \(w_T \ge \lambda\), we obtain that
    \[
        \mathrm{Var}[B'_T] = \left(1- \frac{p^{w_T - \lambda + 1}}{1 + p}\right)\frac{p^{w_T- \lambda + 1}}{1+p}.
    \]
    Similar to the bias, the variance is maximized in the boundary cases, which proves the claim.
\end{proof}

Lastly, we bound the covariance between \(B'_T\) and \(B'_{T'}\), which rely on the same noisy weight information.

\begin{figure}
\centering
\begin{tikzpicture}[
        vertex/.style={draw, circle, inner sep=2pt},
        scale=1
    ]
        \node[vertex,fill=blue!20] (v) at (0, 0) {\(v\)};
        \node[vertex,fill=orange!20] (u) at (4, 0) {\(u\)};
        \node[vertex] (x) at (2, 1) {\(x\)};
        \node[vertex] (y) at (2, -1) {\(y\)};

        \draw (v) -- (y);
        \draw (v) -- (x);
        \draw (u) -- (y);
        \draw (u) -- (x);
        \draw[->] (v) -- (1, 0);
        \draw[->] (u) -- (3, 0);
        \draw[blue, very thick] (x) -- (y);

        \node[vertex, fill=blue!20] at (5, 1) {};
        \node[vertex, fill=orange!20] at (5, 0) {};
        \node[align=center, text width=3cm] at (6.5, 1) {responsible for \(T=\{v,x,y\}\)};
        \node[align=center, text width=3cm] at (6.5, 0) {responsible for \(T' = \{u,x,y\}\)};

        \draw[very thick, blue] (4.75, -1) to (5.25, -1);
        \node[align=center, text width=2cm] at (6.5, -1) {noisy weight};
    \end{tikzpicture}
    \caption{Visualization of \(C'_4\) instance introducing covariance}
     \label{fig:cov_ins}
\end{figure}

\begin{lemma}\label{lem:bias_cov}
    For the triangles \(T \in \Delta_v\) and \(T' \in \Delta_{v'}\) that use the same noisy weight information the covariance of the estimators \(B'_T\) and \(B'_{T'}\) is bounded by 
        \(2\frac{e^{-\varepsilon_1}}{1 + e^{-\varepsilon_1}}\).
\end{lemma}
\begin{proof}
    Let \(T,T' \in \Delta\) denote triangles that share the same noisy weight information, \(B'_T, B'_{T'}\) be the respective biased estimators and \(Z \sim DLap(p)\) be the noise added to the weight information.
    
    Both estimator are \(1\) at the same time when \(\max(w_T , w_{T'}) + Z < \lambda\).
    This results in the following formula for the covariance:
    \[
        \Pr[Z < \lambda - \max(w_T, w_{T'})] - \Pr[Z < \lambda - w_T]\Pr[Z < \lambda - w_{T'}].
    \]
    We make a case distinction on the values of \(w_T\) and \(w_{T'}\) and use the probabilities from Lemma \ref{lem:bias_exp} to bound the covariance in each case. \\
    \textbf{Case \(w_T < \lambda\) and \(w_{T'} < \lambda\):}
    Using the same techniques as in Lemma~\ref{lem:bias_exp} gives:
    \[
        \Pr[Z < \lambda - \max(w_T, w_{T'})] = 1- \frac{p^{\lambda - \max(w_T, w_{T'})}}{1+p},
    \]
    which results in a covariance of
    \[
        \frac{1}{1+p} \cdot \left(p^{\lambda - w_T} + p^{\lambda - w_{T'}} - p^{\lambda -\max(w_T, w_{T'})} - \frac{p^{2\lambda - w_T - w_{T'}}}{1+p}\right).
    \]
This covariance is maximized when \(w_T = \lambda - 1\) and \(w_{T'} = \lambda - 1\). In this case, the covariance of \(B'_T\) and \(B'_{T'}\) is not larger than
    \(\frac{2p}{1+p}.\)

    For the remaining cases, we provide only the resulting covariance and the corresponding upper bound obtained at the boundary of each case.\\
    \textbf{Case \(w_T < \lambda\) and \(w_{T'} \ge \lambda\):}
        \(\frac{p^{w_{T'}- w_T + 1}}{(1+p)^2} \le \frac{p}{1+p}.\) \\
    \textbf{Case \(w_T \ge \lambda\) and \(w_{T'} < \lambda\):}
        \(\frac{p^{w_{T}- w_{T'} + 1}}{(1+p)^2} \le \frac{p}{1+p}.\) \\
    \textbf{Case \(w_T \ge \lambda\) and \(w_{T'} \ge \lambda\):}
    \[
        1/(1+p)\left(p^{\max(w_T, w_{T'})- \lambda + 1} - \frac{p^{w_T + w_{T'} - 2\lambda + 2}}{1+p}\right) \le \frac{p}{1+p}.
    \]
\end{proof}

This lets us bound the total variance of the biased estimators.
\begin{lemma}\label{lem:bias_total_var}
    Let \(\#C'_4\) denote the number of \(C_4'\) instances in our assignment \(\rho\). 
    The variance \(\mathrm{Var}\left[\sum_{v \in V} f'_v(w^v)\right]\) is bounded by \(O\left(\frac{|\Delta|}{\varepsilon_1} + \frac{\#C'_4}{\varepsilon_1}\right)\).
\end{lemma}
\begin{proof}
    The result follows from the bounds introduced in Lemmas \ref{lem:bias_var} and \ref{lem:bias_cov}.
    \[
    \begin{split}
        &\sum_{v \in V} \mathrm{Var}[f_v'(w^v)] + \sum_{u,x \in V} \sum_{v,v'\in N_{u,x}} \mathrm{Cov}[f_v'(w^v), f'_{v'}(w^{v'})] \\
        &\le O\left(\frac{1}{\varepsilon_1}\right)\sum_{v \in V} |\Delta_v| + \#C'_4 \frac{2e^{-\varepsilon_1}}{ 1+ e^{-\varepsilon_1}} 
        = O\left(\frac{|\Delta|}{\varepsilon_1} + \frac{\#C'_4}{\varepsilon_1}\right).
    \end{split}
    \]
\end{proof}

Lastly, we bound the variance introduced by publishing the local triangle counts in the final step.

\begin{lemma}\label{lem:lap_var}
    The variance of the Laplace query in the second step is bounded by 
        \(O\left(\frac{d_{max}m}{\varepsilon^2_2}\right).\)
\end{lemma}
\begin{proof}
The variances of publishing the local counts are independent from each other and the variance of the Laplace query \(Lap(b)\) is equal to \(2b^2\).   
Hence, the Laplace query for publishing the local counts in Step 2 is bounded by
        \(2\sum_{v \in V}\frac{GS(f'_v)^2}{\varepsilon^2_2} \le 4\frac{d_{max}}{\varepsilon^2_2}m\).
    The inequality follows from the fact that a single incident edge of \(v\) can be in at most \(d_v - 1\) triangles. 
    The sensitivity \(GS(f'_v)\) is therefore bounded by the degree \(d_v\). This then also results in \(\sum_{v \in V} GS(f'_v) \le \sum_{v \in V} d_v = 2m\).
\end{proof}

We are now ready to proof the bound on the expected squared \(\ell_2\)-error of Theorem~\ref{thm:bias_err}.
\begin{proof}[Proof of Theorem~\ref{thm:bias_err}]
    This follows from Lemmas \ref{lem:bias_squared}, \ref{lem:bias_total_var} and \ref{lem:lap_var}.
\end{proof}

We note that in the high-privacy regime (\(\varepsilon \to 0\)) the terms depending on \(\varepsilon_1\) in the covariance and variance are bounded by some constant. As explained in Section~\ref{sec:analysis}, the lower dependency on the privacy budget is the main advantage of the biased variant compared to the unbiased one.

\subsection{Precision of the unbiased estimator}\label{subsec:unb_prec}
In this section, we analyze Algorithm \ref{alg:basis} when the unbiased estimator \(U'_T\) is employed at Line 7. 
The analysis follows the steps of the previous section.

\begin{lemma}
    For a triangle \(T \in \Delta\), the estimator \(U'_T\) is unbiased, i.e. 
        \(\mathbb{E}[U'_T] = 1\{w_T < \lambda\}\).
\end{lemma}
\begin{proof}
The claim follows from Lemma 10 in \cite{unbiased_histo}. 
Let \(f\) be the function defined in that lemma. Then,
\[
\mathbb{E}[h(w'_T)]
= 1 - \mathbb{E}[f(w'_T - \lambda)]
= 1 - [w_T \ge \lambda]
= \mathbf{1}[w_T < \lambda].
\]
We also note that the same proof applies for all \(w_T, \lambda \in \mathbb{Z}\), and is not limited to \(\mathbb{N} \cup \{0\}\) as originally stated in \cite{unbiased_histo}.
\end{proof}

Hence, the expected squared \(\ell_2\)-biased consists only of the variance introduced by the estimator and the variance of the general Laplace query for publishing the local below-threshold counts.

\begin{lemma}
    For a triangle \(T \in \Delta\), the variance of the estimator \(U'_T\) is given by 
        \(4p^{|w_T - \lambda| + 1}\left(\frac{p}{(1-p)^3}+ 1-p\right).\)
\end{lemma}
\begin{proof}
    This follows from Theorem 10 in \cite{unbiased_histo}.
\end{proof}

\begin{lemma}\label{lem:unb_var}
    For a node \(v \in V\), the variance  \(\mathrm{Var}[f'_v(w^v)]\) is bounded by 
        \(O\left(\frac{|\Delta_v|}{\varepsilon^3_1}\right)\).
\end{lemma}
\begin{proof}
The variance of an individual estimator \(U'_T\) with \(T \in \Delta_v\) is maximized in the boundary case \(\lvert w_T - \lambda \rvert = 0\).
Moreover, each triangle counted by \(v\) uses an independent noisy weight, so there is no covariance between different estimators. Hence, 
    \[
        \mathrm{Var}[f'_v(w^v)] \le |\Delta_v|4p\left(\frac{p}{(1-p)^3} + 1 -p\right) = O\left(\frac{|\Delta_v|}{\varepsilon^3_1}\right).
    \]
\end{proof}

As discussed in Appendix \ref{sec:bias_ana}, any two estimators \(U'_T\) and \(U'_{T'}\) that rely on the same noisy weight share a nonzero covariance.
To bound the covariance contribution of each \(C'_4\) instance, we apply the Cauchy--Schwarz inequality.
\begin{lemma}\label{lem:unb_cov}
    For two estimators \(U'_T, U'_{T'}\) that use the same noisy weight information, the covariance is bounded by 
        \[4p\left(\frac{p}{(1-p)^3} + 1 -p\right) = O\left(\frac{1}{\varepsilon^3_1}\right).\]
\end{lemma}
\begin{proof}
We bound the covariance of the estimators \(U'_T\) and \(U'_{T'}\) using the Cauchy--Schwarz inequality:
\[
4\left(\frac{p}{(1-p)^3} + (1-p)\right)\sqrt{p^{\,|w_T - \lambda| + |w_{T'} - \lambda| + 2}}.
\]
This bound is maximized when both triangle weights are exactly at the threshold, i.e., \(w_T = \lambda\) and \(w_{T'} = \lambda\). 
\end{proof}

We then obtain the variance introduced by the unbiased estimator in the following theorem.
\begin{lemma}\label{lem:unb_total_var}
    The variance introduced by using the unbiased estimator with noisy edge weights is bounded by
        \(O\left(\frac{|\Delta|}{\varepsilon^3_1} + \frac{\#C'_4}{\varepsilon^3_1}\right)\)
\end{lemma}
\begin{proof}
    This follows from Lemmas \ref{lem:unb_var} and \ref{lem:unb_cov}.
\end{proof}

Before presenting the final theorem on the expected \(\ell_2\)-error of the algorithm with the unbiased estimator, we first bound the variance arising from releasing the local counts in Step~2.
We note that the unbiased estimator depends on the privacy budget \(\varepsilon_1\). As a consequence, both its sensitivity and the parameter of the Laplace query used in Step~2 also depend on \(\varepsilon_1\).

\begin{lemma}\label{lem:unb_lap}
    The variance of the Laplace query in the second step of algorithm \ref{alg:basis} with unbiased estimator is 
        \(O\left(\frac{d_{max}}{\varepsilon^4_1\varepsilon^2_2} m\right)\).
\end{lemma}
\begin{proof} We obtain that
    \[
    \begin{split}
        2\sum_{v \in V}\frac{GS(f'_v)^2}{\varepsilon^2_2} & \le \frac{4}{\varepsilon^2_2}\left(1+ 2\frac{p}{(1-p)^2}\right)^2d_{max}m = O\left(\frac{d_{max}}{\varepsilon^4_1\varepsilon^2_2} m\right).
    \end{split}
    \]
\end{proof}

\begin{proof}[Proof of Theorem~\ref{thm:final-unbiased}]
    This follows from Lemmas \ref{lem:unb_total_var} and \ref{lem:unb_lap}.
\end{proof}

\section{Assignment function} \label{appendix:assignment}

In this section we provide the algorithm for computing the optimal assignment and the proofs for the lemmas and theorems of Section~\ref{sec:mincov}.

\subsection{Computation of optimal assignment}
Given a graph \(G=(V,E)\) with the set of triangles \(\Delta\), we construct a min-cost max-flow network to compute the optimal assignment function. 
In the following, \(N=(V_N, E_N)\) denotes the min-cost max-flow network. 

\subsubsection{Network construction}
In addition to a source \(s\) and target \(t\), we introduce a node \(v_T\) for every triangle \(T \in \Delta\) and a node \(v_e\) for edge \(e \in E\): 
\[
    V_N = \left\{ s,t \right\} \cup \left\{v_T : T \in \Delta\right\} \cup \left\{v_e : e \in E \right\}
\]
For an edge \((u,v) \in E_N\), \(a(u,v) > 0\) denotes the capacity and \(b(u,v)\) denotes the cost for sending a unit of flow along the edge. We represent the edge including the capacity and cost directly as a tuple of the form \((u, v, a(u,v), b(u,v))\).
We first introduce edges linking the source \(s\) to the triangle nodes \(v_T\) with capacity \(1\) and costs \(0\).
\[
    \left\{(s,v_T, 1, 0) : \forall T \in \Delta \right\}
\]
Secondly, a triangle node \(v_T\) is connected to all edge nodes \(v_e\) where \(e\) is contained in the triangle \(T\) with capacity \(1\) and cost \(0\).
\[
    \left\{(v_T, v_e, 1, 0) : \forall e \in T, T \in \Delta\right\}
\]
Let \(d_{in}(v_e)\) denote the incoming degree of an edge node \(v_e\) in \(N\). This represents the number of triangles the edge \(e\) is contained in.
We now add \(d_{in}(v_e)\) edges from the edge node \(v_e\) to the target \(t\) with capacity \(1\) and costs in \(\{0, ..., d_{in}(v_e)-1\}\).
\[
    \left\{(v_e, t, 1, b) : b \in \{0,..., d_{in}(v_e)-1\}, \forall e \in E  \right\}
\]
 
\subsubsection{Assignment extraction}
We now describe how to extract the assignment \(\rho\) alongside the costs of the assignment \(c(\tau)\) for a max-flow min-cost solution of the network \(N\).
Let \(f(u, v)\) denote the flow of an edge \((u,v) \in E_N\) in the network.
For a triangle \(T = \{u,v,w\} \in \Delta\), let \((v_T, v_e) \in E_N\) be a network edge with flow one: \(f(v_T, v_e) = 1\).
We then assign \(T\) to the node that is not included in the edge \(e \in E\).
For \(e = \{u,v\}\), this results in \(\rho(T) = w\).

We first remind the reader of the min-cost max-flow integrality property.
\begin{lemma}[\cite{ahuja1988network}]
    Any minimum cost network flow problem instance whose demands, supplies and capacities are all integers has an optimal solution with integer flow on each edge. Such an optimal solution can be found in 
\(O\bigl(F \cdot (|E_N| + |V_N|\log |V_N|)\bigr)\) time for an input network with \(|V_N|\) nodes, \(|E_N|\) edges, and maximum flow value \(F\). \label{lem:max-flow}
\end{lemma}

We observe that, in the flow obtained in Lemma~\ref{lem:max-flow}, for each \(T \in \Delta\), exactly one edge in
\(\{(v_T, v_e, 1, 0) : e \in T\}\)
has flow value one. We define the assignment \(\rho_f\) by setting \(\rho_f(T)=e\) whenever the edge \((v_T, v_e, 1, 0)\) carries one unit of flow. In what follows, we assume that \(f\) is an optimal integer-flow solution.
\begin{lemma}\label{lem:flow2ass}
If the flow solution \(f\) has cost \(C\), then \(c(\rho_f)=C\).
\end{lemma}
\begin{proof}
    We first show that for each triangle \(T = \{e_1, e_2, e_3\} \in \Delta\) there exist exactly one edge \((v_T, v_{e_i}) \in E_N\) with \(i \in \{1,2,3\}\) and flow \(1\). 
    Given that the max-flow value is \(|\Delta|\) and each source edge \((s, v_T)\) has capacity \(1\), every such edge \((s, v_T)\) is saturated with flow \(1\). By flow conservation at \(v_T\) and flow integrality, exactly one outgoing edge \((v_T, v_{e_i})\) has a flow of \(1\).

    We now proof that the costs of the flow solution is equivalent to the assignment costs. The network has only costs on the edges of the form \((v_e, t)\). Let \(v_e\) be an edge node with incoming flow \(f_{in}(v_e)\). The costs are minimized by satisfying the outgoing edges in increasing order of costs. The total costs by flow traveling through \(v_e\) is then \[\sum_{i = 0}^{f_{in}(v_e)-1} i = \binom{f_{in}(v_e)}{2}.\]
    Since the edges from \(v_e\) to \(t\) and \(v_{e'}\) to \(t\) are distinct, the total costs is given by \(\sum_{e \in E} \binom{f_{in}(v_e)}{2}\). An edge \(e \in E\) is assigned to \(T\) if \(f(v_T,v_e) = 1\), so \(f_{in}(v_e)\) represents exactly the load \(l(e)\).
\end{proof}

\begin{lemma}\label{lem:ass2flow}
    Let \(\rho\) denote an assignment for graph \(G\) with cost \(C\), then there exists a min-cost max-flow solution \(f\) for the network \(N\) with flow \(|\Delta|\) and costs \(C\).
\end{lemma}
\begin{proof}
    Let \(\rho\) be an assignment for the graph \(G\) with cost \(C\). We show how to construct a solution \(f\). Firstly, for each source-triangle edge \((s, v_T)\) set the flow to \(1\). 
    For \(T = \{u,v,w\}\) with \(\rho(T) = v\), set \(f(v_T, v_{\{u,w\}}) = 1\).
    Let \(f_{in}(v_e)\) denote the incoming flow of \(v_e\), choose \(f_{in}(v_e)\) outgoing edges to \(t\) in order of increasing costs.
    
    We first show that the constructed solution is valid.
    The capacity constrained is satisfied since the set flow is bounded by \(1\) and the capacity is \(1\) for every edge in \(N\). 
    Flow conservation is given on triangle nodes since each has incoming flow of \(1\) and also outgoing flow of \(1\) by the fact that each triangle is assigned to exactly \(1\) edge in the assignment \(\rho\). 
    For edge nodes \(v_e \in V_N\) flow conservation is satisfied by the construction of the solution. 
    Lastly, the outgoing flow of \(s\) is \(|\Delta|\) and the incoming flow of \(t\) is also \(|\Delta|\). 
    The cost of the flow is then given by
    \[
        \sum_{e \in E} \binom{f_{in}(v_e)}{2} = \sum_{e \in E_G} \binom{l(e)}{2}
    \]
\end{proof}

\begin{theorem}
    Let \(f\) denote the optimal min-cost max-flow solution to the network \(N\), then the extracted assignment \(\rho_f\) is an optimal assignment. 
\end{theorem}
\begin{proof}
    Follows from Lemma~\ref{lem:flow2ass} and Lemma~\ref{lem:ass2flow}.
\end{proof}

\subsubsection{Running time analysis}
The constructed network allows us to apply any min-cost max-flow algorithm to compute the optimal flow solution. 
In the following, we show the size of the flow-network \(N\) which provides the bounds on the running time. We assume that the network was converted to a simple graph without multi-edges by introducing stub nodes.

\begin{lemma}\label{lem:flow_net_nodes}
    The number of nodes \(|V_N|\) is \(2 + 4|\Delta| + |E|\).
\end{lemma}
\begin{proof}
    The number of nodes without stubs follows directly from the definition.
    \[
        |\left\{ s,t \right\} \cup \left\{v_T : T \in \Delta\right\} \cup \left\{v_e : e \in E \right\}| = 2 + |\Delta| + |E| 
    \]
    A stub is generated for the multi-edges \((v_e,t)\). By the construction \(d_{in}(v_e)\) multi-edges between \(v_e\) and \(t\) exist. Since the total number of incoming edges on the edge-node level is equal to the number of outgoing edges on the triangle-node level and each triangle-node \(v_T\) has 3 outgoing edges this results in the following total amount for the number of stub nodes:
    \(
        \sum_{e \in E} d_{in}(v_e) =  3|\Delta|
    \)
\end{proof}

\begin{lemma}\label{lem:flow_net_edges}
    The number of edges \(|E_N|\) is \(10|\Delta|\).
\end{lemma}
\begin{proof}
    There exist \(|\Delta|\) edges between the source \(s\) and the triangle-nodes, \(3|\Delta|\) edges between the triangle-nodes and edge-nodes, and \(6|\Delta|\) edges between edge-nodes and the target. The latter follows from the fact that there exist \(3|\Delta|\) stubs and each is incident to exactly two edges.
\end{proof}

We are now able to provide the proof of Theorem~\ref{thm:assgn_time}.

\begin{proof}[Proof of Theorem~\ref{thm:assgn_time}]
    To compute the min-cost max-flow, we utilize the Successive Shortest Path algorithm. For a network with a known maximum flow \(F\), node set \(V_N\), and edge set \(E_N\), each augmentation step requires a shortest-path computation which takes \(O(|E_N| + |V_N| \log |V_N|)\) time. The total running time is therefore bounded by \(O(F \cdot (|E_N| + |V_N| \log |V_N|))\). 
    
    From Lemma~\ref{lem:flow_net_nodes} and Lemma~\ref{lem:flow_net_edges}, the network size is bounded by \(|V_N| = O(|\Delta| + |E|)\) and \(|E_N| = O(|\Delta|)\). By construction, the maximum flow is exactly \(F = |\Delta|\). 
    Substituting these values into the algorithmic complexity, we obtain:
    \[
        O(|\Delta| \cdot (|\Delta| + (|\Delta| + |E|) \log(|\Delta| + |E|)))
    \]
    Expanding this expression yields the final theoretical bound on the running time:
    \[
        O(|\Delta|^2 + |\Delta|(|\Delta| + |E|) \log(|\Delta| + |E|))
    \]
\end{proof}

\subsection{Greedy algorithm for the computation of the optimal assignment}

Algorithm~\ref{alg:greedy_c} presents the greedy approach described in Section~\ref{sec:mincov}, whose running time is linear in \(|\Delta|\).

\begin{algorithm}
    \caption{\textsc{Greedy algorithm for computing assignment \(\rho\)}}\label{alg:greedy_c}
    \begin{algorithmic}[1]
        \State Initialize \(l(e) = 0 \quad \forall e \in E\)
        \For{\(T \in \Delta\)}
            \State \(e = \argmin_{e \in T} l(e)\)
            \State Let \(v\) be the only node in \(T / e\). Assign \(\rho(T) = v\) 
            \State \(l(e) = l(e) + 1\)
        \EndFor
    \end{algorithmic}
\end{algorithm}

We restate the optimization problem introduced in Section~\ref{sec:mincov}. As previously shown, solving the optimization problem is equivalent to finding the assignment with minimal costs.

For a triangle \(T \in \Delta\) and edge \(e \in T\), let \(x_{T, e} \in \{0,1\}\) denote if \(T\) is assigned to \(e\).
\begin{equation}\label{eq:opt_def}
\begin{aligned}
\min \quad & \sum_{e \in E} (\sum_{T \in \Delta} x_{T,e})^2\\
\textrm{s.t.} \quad & \sum_{e \in T} x_{T, e} = 1 & \quad \forall T \in \Delta\\
  &x_{T,e} \in \{0,1\}  &\quad \forall T \in \Delta, e \in T \\
\end{aligned}
\end{equation}

\begin{lemma}\label{thm:opt_load_def}
The optimization problem (\ref{eq:opt_def}) reduces to minimizing the \(\ell_2\) norm in the classic offline load-balancing problem, where unit-size jobs must be assigned to machines under restricted assignment constraints.
\end{lemma}
\begin{proof}
    Let a triangle \(T \in \Delta\) represent job \(J_T\) and let edge \(e \in E\) represent machines \(M_e\). A job \(J_T\) can then be scheduled to exactly one of the three machines \(M_{e_i}, M_{e_j}, M_{e_k}\) such that \(e_i, e_j, e_k \in T\). 
    Each job increases the load on its assigned machine by 1. The squared \(\ell_2\)-norm of the machine loads represents the objective value in problem (\ref{eq:opt_def}).
\end{proof}

\begin{lemma}\label{lem:greedy_part_approx}
    Algorithm \ref{alg:greedy_c} has an approximation ratio of \(3+ 2\sqrt{2}\) for the optimization problem (\ref{eq:opt_def}).
\end{lemma}
\begin{proof}
In \cite{greedy_c}, it was shown that the greedy algorithm for the offline load-balancing problem attains a \(1 + \sqrt{2}\) approximation ratio for the \(\ell_2\) norm. This implies an approximation ratio of \(3 + 2\sqrt{2}\) for the squared \(\ell_2\) norm.
\end{proof}

We are now able to proof the approximation ratio as stated in Theorem~\ref{thm:greedy_approx}.

\begin{proof}[Proof of Theorem~\ref{thm:greedy_approx}]
    Let \(l(e)\) denote the loads in the greedy assignment, and \(l^*(e)\) denote the loads in the optimal assignment. From the definition of the assignment function, the number of triangles is conserved, meaning \(\sum_{e \in E} l(e) = \sum_{e \in E} l^*(e) = |\Delta|\).
    
    The cost of the optimal assignment is: \(\#C'_{4, OPT} = \frac{1}{2}\sum_{e \in E} (l^*(e))^2 - \frac{1}{2}|\Delta| \).
    Rearranging this equation to isolate the sum of squared loads gives:
    \[ \sum_{e \in E} (l^*(e))^2 = 2\#C'_{4, OPT} + |\Delta| \]
    Similarly, for the greedy algorithm, we have:
    \[ \sum_{e \in E} (l(e))^2 = 2\#C'_{4, ALG} + |\Delta| \]
    
    By Lemma \ref{lem:greedy_part_approx}, we know the greedy algorithm achieves a \(3+2\sqrt{2}\) approximation for the sum of squared loads:
    \[ \sum_{e \in E} (l(e))^2 \le (3+2\sqrt{2})\sum_{e \in E} (l^*(e))^2 \]
    
    Substituting the sum of squared loads and rearranging the inequality yields the statement.
\end{proof}

\subsection{Bounds for planar graphs}

We now present the proofs for the upper and lower bounds of the precision on planar graphs of Section~\ref{sec:mincov}. We denote the number of nodes by \(n\) and the number of edges by \(m\) in this section.
We first state the required definitions for this section.

\begin{definition}[Planar graph]
    A graph \(G = (V, E)\) is \emph{planar} if it admits a drawing in the plane such that edges intersect only at their common endpoints.
\end{definition}

\begin{definition}[Maximal planar graph]
    A planar graph is \emph{maximal planar} (or a \emph{triangulation}) if no edge can be added without violating planarity, or equivalently, if every face of any planar embedding, including the outer face, is a triangle.
\end{definition}

\begin{lemma}\label{lem:max_plan_tri}
    For any maximal planar graph \(G\) with \(|\Delta| \ge 1\),
    \(
        |\Delta| \le m-2.
    \)
\end{lemma}
\begin{proof}
    From \cite{hakimi1979number}, it is known that any maximal planar graph with \(n \ge 4\) has \(|\Delta| \le 3n-8\). Moreover, the number of edges is \(m =3n-6\) in maximal planar graphs. This shows the \(|\Delta| \le m-2\) statement for \(n \ge 4\). For \(n=3\), there exists only the graph consisting of a single triangle for which the statement holds trivially. 
\end{proof}

We now generalize Lemma~\ref{lem:max_plan_tri} for general planar graphs.
\begin{lemma}\label{lem:plan_tri_bound}
    For any planar graph with \(|\Delta| \ge 1\), \(|\Delta| \le m-2\).
\end{lemma}
\begin{proof}
    Let \(G = (V,E)\) be a planar graph with \(|\Delta| \ge 1\).
    If the graph is maximal planar, then the statement follows from Lemma~\ref{lem:max_plan_tri}. 
    
    In the following, assume that \(G\) is not maximal planar. We generate a maximal planar graph \(G' = (V, E')\) by adding edges to \(G\) without violating planarity. 

    Since \(G\) is not maximal planar, there exists at least one face \(\mathsf{f}\) that is bounded by \(k \ge 4\) vertices. Treating the boundary of \(\mathsf{f}\) as a simple polygon, we can triangulate its interior. As proven by \cite{garey1978triangulating}, any simple polygon of \(k\) vertices can be fully partitioned into exactly \(k-2\) triangles by adding exactly \(k-3\) non-intersecting internal chords. 

    Let \(\mathsf{c_f}\) denote the number of edges (chords) added to triangulate face \(\mathsf{f}\), and \(\mathsf{d_f}\) denote the number of new triangles created. Based on \cite{garey1978triangulating}: \(\mathsf{c_f} = k - 3\) and \(\mathsf{d_f} \ge k - 2\). 
    We note that additional triangles might be generated outside the face resulting an inequality for \(\mathsf{d_f}\).
    It is guaranteed that for every face triangulated, the number of new triangles generated is strictly greater than the number of edges added, \(\mathsf{d_f > c_f}\).

    We repeat this triangulation process for all non-triangular faces until the resulting graph \(G'\) is maximal planar. 
    Let \(m',\Delta'\) denote the number of edges and the set of triangles after the process, and let \(\mathsf{c}\) denote the total number of edges introduced during the entire procedure (\(\mathsf{c} = m' - m\)), and let \(\mathsf{d}\) denote the total number of new triangles introduced (\(\mathsf{d} = |\Delta'| - |\Delta|\)). 

    Because \(G\) was not maximal planar, we must have triangulated at least one face, meaning \(\mathsf{d_f > c_f}\) occurred at least once. 
    Therefore, summing the newly added edges and triangles across the entire graph yields: \(\mathsf{d \ge c + 1}\).

    Since the resulting graph \(G'\) is maximal planar, we apply Lemma~\ref{lem:max_plan_tri} to get the following inequality:
    \[|\Delta'| \le m' - 2\]
    Substituting \(|\Delta'| = |\Delta| + \mathsf{d}\) and \(m' = m + \mathsf{c}\) into the inequality yields:
    \[|\Delta| + \mathsf{d} \le m + \mathsf{c} - 2\]
    Rearranging the terms to isolate \(|\Delta|\) gives:
    \[|\Delta| \le m - 2 - (\mathsf{d - c})\]
    Because we have established that \(\mathsf{d - c} \ge 0\), the term \(-(\mathsf{d-c})\) is less than or equal to zero. Therefore, \(|\Delta| \le m - 2\).    
\end{proof}

We are now ready to proof the existence of an assignment \(\rho\) with \(c(\rho) = 0\) for any planar graph.

\begin{proof}[Proof of Lemma~\ref{lem:planar_opt}]
    Given a planar graph \(G = (V,E)\), we construct the bipartite graph \(H = (A \cup B, E_H)\) with \(A = \left\{v_T : T \in \Delta \right\}, B = \left\{v_e: e \in E\right\}\), and \(E_H = \left\{ \{v_T, v_e\} : \forall e \in T, T \in \Delta\right\}\).
    Let \(\mathcal{N}(v_T) = \left\{v_e : \{v_T, v_e\} \in E_H\right\}\) denote the neighborhood of a triangle node \(v_T \in A\). Let \(\mathcal{N}(S) = \bigcup_{v_T \in S} \mathcal{N}(v_T)\) be the neighborhood of the set \(S \subseteq A\).
    By Hall's theorem there exists a perfect matching on the triangles if for every subset \(S \subseteq A\), \(|S| \le |\mathcal{N}(S)|\). Let \(G(S) = (V, E_S)\) be the subgraph restricted to edges contained in a triangle in \(S\):
    \[
        E_S = \left\{e \in E: \exists v_T \in S \; s.t. \; e \in T\right\}
    \]
    Let \(X_1, ..., X_j\) denote the connected components of \(G(S)\) and let \(S(X_i)\) denote the set of triangles in \(X_i\). By construction every triangle in \(S\) is in exactly one connected component \(X_i\): \(\sum_{X_i} |S(X_i)| = |S|\). 
    Moreover, each connected component contains at least one triangle: \(|S(X_i)| \ge 1\).
    
    We can now apply Lemma~\ref{lem:plan_tri_bound} to each component: \(|S(X_i)| \le |E_{X_i}| - 2\). Summing over all connected components, we get:
    \[
        |E_S| = \sum_{X_i} |E_{X_i}| \ge \sum_{X_i} \left( |S(X_i)| + 2 \right) = |S| + 2j \ge |S|
    \]
    Since \(|\mathcal{N}(S)| = |E_S|\), by Hall's theorem, there exist a perfect triangle-side matching and therefore a triangle assignment \(\rho\) with \(c(\rho) = 0\).  
    For planar graphs \(|\Delta| = O(n)\) and \(m = O(n)\), the \(O(n^2\log n)\) running time follows from Theorem~\ref{thm:assgn_time}.
\end{proof}

In the following we provide the construction of the \(\Omega(n^2)\) lower bound on the expected squared \(\ell_2\)-error of any \(\varepsilon\)-LWDP algorithm.
\begin{proof}[Proof of Lemma~\ref{lem:planar_lower}]
    Let \(G=(V,E,w)\) be a graph with \(V=\{v_1,\dots,v_n\}\) consisting of \(n-2\) triangles that share the common edge \(\{v_1,v_2\}\). That is,
\[
    E=\{\{v_1,v_2\}\}
    \cup
    \left\{\{v_1,v_j\},\{v_2,v_j\}: v_j\in V\setminus\{v_1,v_2\}\right\}.
\]
Let the weights of all edges except \(\{v_1,v_2\}\) be \(0\), and let the threshold be \(\lambda=1\). We further assume that the weights of all edges except \(\{v_1,v_2\}\) are public information, while the weight of \(\{v_1,v_2\}\) belongs to \(\{0,1\}\).

Suppose, for contradiction, that for every constant \(c(\varepsilon)\) depending on \(\varepsilon\) such that \(c(\varepsilon) > 0\), there exists an \(\varepsilon\)-LWDP algorithm \(\mathcal{M}\) whose expected squared \(\ell_2\)-error is smaller than \(c(\varepsilon) n^2 \).

Let \(\mathcal{G}\) be an algorithm that guesses the weight of \(\{v_1,v_2\}\) from the output of \(\mathcal{M}\). Specifically, \(\mathcal{G}\) outputs \(0\) if the output of \(\mathcal{M}\) is smaller than \((n-2)/2\), and outputs \(1\) otherwise. By the post-processing property of \(\varepsilon\)-LWDP, the algorithm \(\mathcal{G}\) is also \(\varepsilon\)-LWDP.

The two graphs corresponding to \(w_{v_1v_2}=0\) and \(w_{v_1v_2}=1\) are neighboring under \(\varepsilon\)-LWDP. Therefore, \(\mathcal{G}\) must guess the weight incorrectly with probability at least
    \(\frac{1}{1+e^{\varepsilon}}\).
Whenever \(\mathcal{G}\) guesses incorrectly, the squared \(\ell_2\)-error of \(\mathcal{M}\) is at least
    \(\frac{(n-2)^2}{4}\).
Hence, the expected squared \(\ell_2\)-error of \(\mathcal{M}\) is at least
    \(\frac{1}{1+e^{\varepsilon}}\cdot \frac{(n-2)^2}{4}\).
This contradicts the assumption that the expected squared \(\ell_2\)-error is smaller than \(c(\varepsilon)  n^2\) for every \(c(\varepsilon)>0\). Therefore, no such \(\varepsilon\)-LWDP algorithm \(\mathcal{M}\) exists.
\end{proof}

\subsection{Bounds for degeneracy-bounded graphs}

In the following, proof the statements regarding graphs with degeneracy \(\gamma\). 

\begin{definition}[Degeneracy]
    For a graph \(G\) the degeneracy \(\gamma(G)\) is the smallest number, such that in any subgraph of \(G\) there exists a node with induced degree at most \(\gamma(G)\).
\end{definition}

For a graph \(G = (V,E)\), let \(\prec\) be an ordering of the nodes. We say \(v\) precedes \(u\) if \(v \prec u\). Let \(d^\prec_v\) denote the forward degree of \(v \in V\): \[
    d^\prec_v = |\left\{\{u,v\} \in E : v \prec u\right\}|.
\]

\begin{lemma}\label{lem:deg_ord}
    Let \(G = (V,E)\) be a graph with ordering \(\prec\). Then there exists a triangle-node assignment \(\rho\) with \(c(\rho) = \sum_{v\in V} (d^\prec_v)^3\).
\end{lemma}
\begin{proof}
    For a triangle \(T = \{u,v,w\} \in \Delta\), assign the triangle to the edge whose nodes occur first in the ordering. \(\rho(T) = v\) if and only if \(u \prec v\), and \(w \prec v\). The load of an edge \(\{u,w\} \in E\) is therefore upper bounded by \(\min\{d^\prec_u, d^\prec_w\}\). 
    \[  
        \#C'_4 \le \sum_{\{u,w\} \in E} \binom{\min\{d^\prec_u, d^\prec_w\}}{2} 
    \]
    The statement then follows:
    \[
        \sum_{\{u,w\} \in E} \binom{\min\{d^\prec_u, d^\prec_w\}}{2} \le \sum_{v \in V} d_v^\prec \binom{d^\prec_v}{2} = O(\sum_{v \in V}(d^\prec_v)^3).
    \]
\end{proof}

\begin{proof}[Proof of Lemma~\ref{lem:deg_ass}]
    Let \(G = (V, E)\) be a graph with degeneracy bounded by \(\gamma\). Let \(\prec\) denote the ordering resulting by repeatedly removing the node with the smallest degree. The forward degree is then bounded by \(\gamma\) and we get \(\#C'_4 = O(n\gamma^3)\) by Lemma~\ref{lem:deg_ord}.
\end{proof}

We now present the proof of Theorem~\ref{thm:deg_lb}. 
We present a proof for the more general framework of \((\varepsilon, \delta)\)-LWDP. 

\begin{definition}[\((\varepsilon, \delta)\)-Local weight differential privacy]
    Let \(\varepsilon > 0\) and \(\delta \in (0,1)\). A randomized query \(\mathcal{R}\) satisfies \((\varepsilon, \delta)\)-local weight differential privacy (\((\varepsilon, \delta)\)-LWDP) if, for any neighboring weight vectors \(w, w'\), and any measurable set of outputs \(S\),
    \[
        \Pr[\mathcal{R}(w) \in S] \leq e^{\varepsilon} \Pr[\mathcal{R}(w') \in S] + \delta.
    \]

    An algorithm \(\mathcal{A}\) is said to be \((\varepsilon, \delta)\)-LWDP if, for any node \(v_i\), and any (possibly adaptive) sequence of queries \(\mathcal{R}_1, \dots, \mathcal{R}_\kappa\) posed to \(v_i\), where each query \(\mathcal{R}_j\) satisfies \((\varepsilon_j, \delta_j)\)-LWDP for \(1 \leq j \leq \kappa\), it holds that
    \(\sum_{j=1}^{\kappa} \varepsilon_j \leq \varepsilon\) and \(\sum_{j=1}^{\kappa} \delta_j \leq \delta.\)
    For \(\delta = 0\), \((\varepsilon, \delta)\)-LWDP is equivalent to \(\varepsilon\)-LWDP.
\end{definition}

As described in Section~\ref{sec:mincov} the proof is based on a reduction from the problem of computing the sum of \(n\) values under local differential privacy.
This proof follows the approach proposed in \cite{eden2025triangle}. They analyzed the setting for general graphs while we provide the analysis for graphs with degeneracy \(\gamma\).

\begin{definition}[\(\textit{SUM}_n\) \cite{eden2025triangle}]
    Given \(x = (x_1, \dots, x_n) \in \{0,1\}^n\), we define \(\textit{SUM}_n\) as
    \[
        \textit{SUM}_n = \sum^n_{i=1} x_i.
    \]
\end{definition}

\begin{lemma}[\cite{chan2012optimal}, \cite{beimel2008distributed}, \cite{joseph2019role}, \cite{eden2025triangle}]\label{lem:sum_bound}
    There exists a constant \(c > 0\) such that for every \(\varepsilon \in (0,1), n \in \mathbb{N}, \alpha_0 \in (0, n]\) and \(\delta \in [0, \frac{1}{10^5} \cdot \frac{\varepsilon^2\alpha^2_0}{n^3\ln{(n^2/\varepsilon\alpha_0)}}]\), if \(\mathcal{A}\) is an \((\varepsilon, \delta)\)-LDP algorithm where each user \(i\) has the private bit \(x_i\) and \(\mathcal{A}\) estimates \(\textit{SUM}_n\) up to an additive error \(\alpha_0\) with probability at least \(2/3\), then \(\alpha_0 \ge c \sqrt{n}/\varepsilon\).
\end{lemma}

In the following, we provide two separate lower bounds that then proof Theorem~\ref{thm:deg_lb}.
\begin{lemma}\label{lem:deg_lb1}
    There exists a family of graphs with degeneracy \(\gamma\) and a constant \(c > 0\) such that for every \(\varepsilon \in (0,1)\), \(n \in \mathbb{N}\), \(\alpha \in (0,  \lfloor \gamma/2 \rfloor(n-\gamma)]\), and
    \[
        \delta \in \left[0, \frac{1}{10^5} \frac{\varepsilon^2\alpha^2}{(n-\gamma)^2k^3\ln\!\left(k^2(n-\gamma)/(\varepsilon \alpha)\right)}\right], \quad \text{where } k = \lfloor \gamma / 2 \rfloor,
    \]
    every \((\varepsilon, \delta)\)-local weight differential privacy algorithm that, given a graph with \(n\) nodes and degeneracy \(\gamma\), outputs an estimate of the number of below-threshold triangles with additive error at most \(\alpha\), with probability at least \(2/3\), must satisfy
    \[
        \alpha \ge c \frac{(n-\gamma)\sqrt{ \lfloor \gamma/2 \rfloor}}{\varepsilon}.
    \]
\end{lemma}
\begin{proof}
    Let \(k = \lfloor \gamma/2 \rfloor\). We reduce from an instance of \(SUM_k\).
    Let \(\vec{x} = (x_1, \dots, x_k) \in \{0,1\}^k\) denote the private bits of the \(k\) users.

    We construct a weighted, undirected graph \(G = (V, E, w)\) with \(n\) nodes as follows. 
    For each \(i \in \{1, \dots, k\}\), introduce two nodes \(u_{2i}\) and \(u_{2i+1}\), together with an edge \(\{u_{2i}, u_{2i+1}\}\) of weight \(-x_i\). 
    Let
    \[
        S = \{u_i : i \in \{2, \dots, 2k+1\}\}
    \]
    denote the set of these nodes. If \(\gamma\) is odd, add one additional node to \(S\) that is not paired with any other node. 
    Let \(H\) be the set of the remaining \(n-\gamma\) hub nodes. For every pair consisting of one node in \(S\) and one node in \(H\), add an edge of weight \(0\).

    We first show that \(G\) has degeneracy \(\gamma\). Let \(A \subseteq V\) and consider the induced subgraph \(G(A)\).
    If \(A \cap H \neq \emptyset\), then any node in \(A \cap H\) has degree at most \(\gamma\) in \(G(A)\), since it is only connected to nodes in \(S\).
    If \(A \cap H = \emptyset\), then \(A \subseteq S\), and every node in \(S\) has degree at most \(1\) in \(G(A)\). 
    Hence, every induced subgraph contains a node of degree at most \(\gamma\), implying that the degeneracy of \(G\) is \(\gamma\).

    Each pair of nodes \(u_{2i}, u_{2i+1}\) is associated with user \(i\), while nodes in \(H\) can be simulated by any user since all incident edge weights are public.

    Let \(\mathcal{A}\) be an \((\varepsilon, \delta)\)-LWDP algorithm that estimates the number of below-threshold triangles with additive error \(\alpha\).
    Fix the threshold \(\lambda = 0\), and let \(o\) denote the output of \(\mathcal{A}\).

    Since there are no edges within \(H\) and only disjoint edges within \(S\), every triangle must consist of two nodes from a pair in \(S\) and one node from \(H\). 
    Thus, all triangles are of the form \(\{u_{2i}, u_{2i+1}, h\}\) for \(i \in \{1, \dots, k\}\) and \(h \in H\).
    The additional node (if \(\gamma\) is odd) does not participate in any triangle.

    If \(x_i = 1\), then the edge \(\{u_{2i}, u_{2i+1}\}\) has weight \(-1\), and each triangle \(\{u_{2i}, u_{2i+1}, h\}\) has total weight \(-1\), which is below the threshold. 
    Hence, user \(i\) contributes exactly \(n-\gamma\) below-threshold triangles. If \(x_i = 0\), no such triangles are below the threshold.

    Therefore,
    \[
        s = \frac{o}{n-\gamma}
    \]
    yields an estimate for \(SUM_k\) with additive error \(\alpha_0 = \frac{\alpha}{n-\gamma}\).

    Substituting \(k = \lfloor \gamma/2 \rfloor\) and \(\alpha_0 = \alpha/(n-\gamma)\) in Lemma~\ref{lem:sum_bound} yields
    \[
        \delta \in \left[0, \frac{1}{10^5} \frac{\varepsilon^2\alpha^2}{(n-\gamma)^2( \lfloor \gamma/2 \rfloor)^3\ln\!\left(( \lfloor \gamma/2 \rfloor)^2(n-\gamma)/(\varepsilon \alpha)\right)}\right]
    \]
    and
    \[
        \alpha \ge c \frac{(n-\gamma)\sqrt{ \lfloor \gamma/2 \rfloor}}{\varepsilon},
    \]
    completing the proof.
\end{proof}
\begin{lemma}\label{lem:deg_lb2}
    There exists a family of graphs with degeneracy \(\gamma\) and a constant \(c > 0\) such that for every \(\varepsilon \in (0,1)\), \(n \in \mathbb{N}\), \(\alpha \in (0, (\gamma-1)\lfloor(n-\gamma)/2 \rfloor]\), and
    \[
        \delta \in \left[0, \frac{1}{10^5}\cdot \frac{\varepsilon^2\alpha^2}{(\gamma-1)^2 k^3 \ln\!\left(\frac{k^2(\gamma-1)}{\varepsilon \alpha}\right)}\right],
        \quad \text{where } k = \left\lfloor \frac{n-\gamma}{2} \right\rfloor,
    \]
    every \((\varepsilon, \delta)\)-local weight differential privacy algorithm that, given a graph with \(n\) nodes and degeneracy \(\gamma\), outputs an estimate of the number of below-threshold triangles with additive error at most \(\alpha\), with probability at least \(2/3\), must satisfy
    \[
        \alpha \ge c \frac{(\gamma-1)\sqrt{k}}{\varepsilon}.
    \]
\end{lemma}

\begin{proof}
    Let \(k = \lfloor (n-\gamma)/2 \rfloor\), and consider an instance of \(SUM_k\) with private bits \(\vec{x} = (x_1,\dots,x_k) \in \{0,1\}^k\).

    Construct a weighted undirected graph \(G=(V,E,w)\) as follows. For each \(i \in \{1,\dots,k\}\), introduce two nodes \(u_{2i}\) and \(u_{2i+1}\) together with an edge \(\{u_{2i},u_{2i+1}\}\) of weight \(-x_i\). Let \(P\) be the set of these \(2k\) nodes. If \(n-\gamma\) is odd, there is one additional node in \(P\) that is not part of any pair; it has no effect on the construction below. Let \(H\) be a set of \(\gamma-1\) hub nodes. Add edges of weight \(0\) between every node in \(P\) and every node in \(H\). Finally, add one isolated node so that the total number of nodes is \(n\).

    We verify that the degeneracy of \(G\) is \(\gamma\). Consider any subset \(A \subseteq V\) and the induced subgraph \(G(A)\). If \(A\) contains a node from \(P\), then that node has degree at most \(\gamma\) in \(G(A)\). If \(A\) contains only nodes from \(H\) and possibly the isolated node, then every node has degree \(0\). Hence every induced subgraph has a node of degree at most \(\gamma\), and since nodes of degree \(\gamma\) exist, the degeneracy is exactly \(\gamma\).

    Fix threshold \(\lambda=0\), and let \(o\) be the output of an \((\varepsilon,\delta)\)-LWDP algorithm \(\mathcal{A}\) that estimates the number of below-threshold triangles with additive error \(\alpha\). Every triangle consists of two nodes from a pair in \(P\) and one node in \(H\). If \(x_i=1\), then the edge \(\{u_{2i},u_{2i+1}\}\) has weight \(-1\), and each triangle \(\{u_{2i},u_{2i+1},h\}\) has total weight \(-1\). 
    Hence, user \(i\) contributes exactly \(|H|=\gamma-1\) below-threshold triangles. If \(x_i=0\), no such triangle is below the threshold.

    Therefore,\(s = \frac{o}{\gamma-1}\)
    is an estimate of \(SUM_k\) with additive error \(\alpha_0 = \frac{\alpha}{\gamma-1}\).

    Substituting \(\alpha_0 = \alpha/(\gamma-1)\) and \(k\) in Lemma~\ref{lem:sum_bound} gives
    \[
        \delta \in \left[0, \frac{1}{10^5}\cdot \frac{\varepsilon^2\alpha^2}{(\gamma-1)^2 k^3 \ln\!\left(\frac{k^2(\gamma-1)}{\varepsilon \alpha}\right)}\right]
    \]
    and
    \[
        \alpha \ge c \frac{(\gamma-1)\sqrt{k}}{\varepsilon},
    \]
    which proves the claim.
\end{proof}

We are now able to provide the complete proof of Theorem~\ref{thm:deg_lb}.
\begin{proof}[Proof of Theorem~\ref{thm:deg_lb}]
    The claim follows from Lemma~\ref{lem:deg_lb1} and Lemma~\ref{lem:deg_lb2}.
\end{proof}

\section{Efficient smooth sensitivity computation for biased estimator}\label{appendix:smooth_bias}

In this section we provide the complete algorithm for the computation of the smooth sensitivity of the function \(f'_v(w^v)\) using the biased estimator \(B'_T\) introduced in Section~\ref{sec:smooth}. 

\subsection{Properties of smooth sensitivity for \(f_v'\)}\label{appendix:smooth_props}

Let \(E(v) = \{e \in E: v \in e\}\) denote the incident edges of \(v\). 
We slightly adapt the notation for \(\beta\)-smooth sensitivity. Instead of maximizing over
\(y \in \mathbb{Z}^d\), we reparameterize using the difference \(z = y - w\) to obtain that: 
 \[
        S^*_{f'_v,\beta}(w) = \max_{z \in \mathbb{Z}^d}(LS_f(w+z)e^{-\beta \cdot |z|})
\]
Let \(B := \{b \in \{-1, 0, 1\}^d : |b|_1 = 1\}\). We have that \(w \sim w'\) if there exists \(b \in B\) such that \(w' = w + b\). Hence, 
\[
    S^*_{f'_v,\beta}(w) = \max_{b \in B}\max_{z \in \mathbb{Z}^d}|f'_v(w +z + b)- f'_v(w + z)|e^{-\beta \cdot |z|}
\]
Assume \(i\) is the unique index with \(b_i \neq 0\).
If the weight of edge \(i\) increases (i.e., \(b_i = 1\)), then the local sensitivity is the number of triangles that contain edge \(i\) and whose total weight is \(\lambda - 1\) in \(w+z\). In \(w +z + b\), each such triangle now has total weight \(\lambda\) and is therefore flipped to being above the threshold \(\lambda\).
Similarly, if the weight of edge \(i\) decreases (i.e., \(b_i = -1\)), then the local sensitivity is the number of triangles that contain edge \(i\) and whose total weight is \(\lambda\) in \(w+z\).

In the following, for a node \(v\), let \(\Delta_v\) denote the set of triangles assigned to \(v\), and for each \(T\in\Delta_v\), let \(T\subseteq E\) denote the set of three edges forming that triangle. 
We analyze the case \(b_i=1\); the case \(b_i=-1\) follows by a symmetric argument.
Define \(\Delta_v(i)=\{\, j\in E(v)\mid \exists\,\{i,j,k\}\in\Delta_v \,\}\) as the set of incident edges that form a triangle together with \(i \in E(v)\). For an edge \(e\), let \(u(e)\) denote the endpoint of \(e\) distinct from \(v\). In this setting,  the local sensitivity is given by the following expression:
\[
    \max_{i \in E(v)}\max_{z \in \mathbb{Z}^d} \sum_{j \in \Delta_v(i)} 1\{w_j + z_j + w'_{u(i)u(j)} = \lambda-1 - w_i - z_i\} e^{-\beta |z|}
\]
We obtain this since \(\lvert f'_v(w+z+b)-f'_v(w+z)\rvert\) equals the number of triangles that cross the threshold when the weight vector changes from \(w+z\) to \(w+z+b\), i.e., triangles that are below the threshold under \(w+z\) but are no less than it under \(w+z+b\). 
Equivalently, it is the number of triangles incident to edge \(i\) whose weight sum under \(w+z\) is exactly \(\lambda-1\), so that adding \(b\) increases the sum to \(\lambda\).

Fix the index \(i\). 
For all \(j \in \Delta_v(i)\), let \(c_j = w_j + w'_{u(i)u(j)}\). 
The smooth sensitivity for a fixed \(i\) and \(b\) becomes,
\begin{equation}\label{eq:smooth_simple}
    \max_{z \in \mathbb{Z}^d} \sum_{j\in \Delta_v(i)}1\{c_j + z_j = \lambda-1-w_i - z_i\}e^{-\beta |z|}
\end{equation}

We aim to solve optimization problem~(2) by choosing \(z\) so that many of the values \(c_j\) are shifted to the \emph{target} of \(t = \lambda - 1 - w_i - z_i\).
Figure \ref{fig:smooth_line_prob} visualizes this setting.
\begin{figure}[h!]
    \centering
    \begin{tikzpicture}
        \draw (-0.5,0) -- (7,0);

        \draw (0.5,0.15) -- (0.5,-0.15) node[below=3pt] {\(c_j\)};
        \draw (3.0,0.15) -- (3.0,-0.15) node[below=3pt] {\(\lambda-1 - w_i -z_i\)};
        \draw (5.5,0.15) -- (5.5,-0.15) node[below=3pt] {\(c_k\)};

        \draw[->, bend left=45, thick, shorten >= 2pt] (0.5, 0) to node[above]{\(z_j\)} (3.0, 0);
        \draw[->, bend right=45, thick, shorten >= 2pt] (5.5, 0) to node[above]{\(z_k\)} (3.0, 0);
    \end{tikzpicture}
    \caption{Representation of smooth sensitivity as a problem of shifting values to a target}
    \label{fig:smooth_line_prob}
\end{figure}

\begin{lemma} \label{thm:target}
There is an optimal solution \(z\) to optimization problem~\ref{eq:smooth_simple} in which \(z_i\) satisfies \(z_i = \lambda - 1 - w_i - c_j\) for some \(j \in \Delta_v(i)\), or \(z_i = 0\). Equivalently, there is an optimal solution whose target \(t\) is either \(t = \lambda - 1 - w_i\) or \(t = c_j\) for some \(j \in \Delta_v(i)\).
\end{lemma}
\begin{proof}
Assume \(z^*\) to be an optimal solution to optimization problem~\ref{eq:smooth_simple} such that \(z_i^* \neq 0\) and \(z_i^* \neq \lambda - 1 - w_i - c_j\) for all \(j \in  \Delta_v(i)\). Let \(t^* = \lambda - 1 - w_i - z_i^*\) be the target in this optimal assignment, let the objective value be \(k^* e^{-\beta \lvert z^* \rvert}\), and let \(s = |\Delta_v(i)|\). This means we have \(k^*\) elements of \(c_j\) shifted to \(\lambda - 1 - w_i - z_i\).

We now either show that all optimal solutions satisfy the claimed property or construct another optimal solution \(z\) satisfying it.
We do so by shifting the target \(t = \lambda - 1 - w_i - z_i\) and adjusting the values \(z_j\) so that the indicators \(1\{ c_j + z_j = \lambda - 1 - w_i - z_i \}\) remain unchanged for all \(j \in \Delta_v(i)\).

   Let \(S_L^{t^*} = \{ j \in [s] \mid c_j \le t^* \ \land\ c_j + z^*_j = t^* \}\) be the set of values that are shifted from the left side of the target \(t^*\), and let \(l = |S_L^{t^*}|\) denote its size:
Similarly, define \(S_R^{t^*} = \{ j \in [s] \mid c_j > t^* \ \land\ c_j + z^*_j = t^* \}\) to be the set of values shifted from the right side of the target \(t^*\), and let \(r = |S_R^{t^*}|\) be its size.

   Consider the case where \(l = r\). Without loss of generality, we assume \(t^* < \lambda -  1 - w_i\). If there is \(j\) such that \(t^* < c_j \leq \lambda - 1 - w_i\), we set \(t\) to smallest \(c_j\) in that range. Otherwise, we set \(t\) to \(\lambda - 1 - w_i\). We then shift all \(k^*\) elements in the sets \(S_L^{t^*}\) and \(S_L^{t^*}\) to the target \(t\).
    Let \(\delta = t - t^*\) be the distance of the target shift. We obtain that:
    \[
        \begin{split}
        |z| &= \sum_{c_j \in S_L^{t}} |z_j| + \sum_{c_j \in S_R^{t}} |z_j| + |z_i| \\
        &\le \sum_{c_j \in S_L^{t^*}} (|z^*_j| + \delta) + \sum_{c_j \in S_R^{t^*}} (|z^*_j| - \delta) + |z_i| \\
        &< |z^*| + (l - r)\delta = |z^*|.
         \end{split}
    \]
    For the second inequality, we use the fact that \(|z_i| < |z^*_i|\).
    From this, we have \(k^*e^{-\beta|z|} > k^*e^{-\beta|z^*|}\). 
    This contradiction shows that for \(l = r\), no such optimal solution \(z^*\) can exist and all optimal solutions must satisfy the property.

    For \(l > r\), we move the target \(t\) to the first \(c_j\) to the left of \(t^*\). Since \(l > r \), such a value must exist. We obtain that:
    \[
        \begin{split}
        |z| &= \sum_{c_j \in S_L^{t}} |z_j| + \sum_{c_j \in S_R^{t}} |z_j| + |z_i| \\
        &\le \sum_{c_j \in S_L^{t^*}} (|z^*_j| - \delta) + \sum_{c_j \in S_R^{t^*}} (|z^*_j| + \delta) + |z_i| \\
        &\le (|z^*| + \delta) + (r - l)\delta \le |z^*|
        \end{split}
    \]
Since \(z\) connects the same values as \(z^*\) and satisfies \(|z| \le |z^*|\), it achieves an objective value that is at least as good as that of \(z^*\).
In the case where \(r < l\), we shift the target \(t\) to the first \(c_j\) located to the right of \(t^*\). Applying the same reasoning as in the previous case confirms that \(z\) is indeed an optimal solution.
\end{proof}

From Lemma \ref{thm:target}, it is sufficient to iterate over \(t \in c \cup \{\lambda -1 - w_i\}\) and find the optimal value of the optimization \ref{eq:smooth_simple} in order to compute the global optimum.
We note that, for a fixed \(t\) and fixed number \(k = \sum_{j=1}^{s}1\{c_j + z_j = \lambda-1-w_i - z_i\}\), it is optimal to shift the \(k\)-closest values in \(c\) to the target \(t\). 
Let \(d_t(j)\) be the \(j\)-th smallest distance from any value in \(c\) to target \(t\) and \(s = |\Delta_v(i)|\). 
The optimal number of connected values \(k^*\) for target \(t\) is then given by
    \[k^* = \argmax_{k \in [s]} k e^{-\beta \sum_{j=1}^{k}d_t(j)}.\]
The idea of our algorithm is to iterate over every possible target \(t\) in non-decreasing order and compute \(k^*\) together with its objective value to compute the global maximum. 
The straightforward approach of iterating over every possible target \(t\) and possible \(k\) results in a running time of \(O(d^2)\) for a fixed edge \(i\).
From now, we show how this running time can be reduced to \(O(d \log^2d)\). For simplicity, we denote \(d_t(j)\) by \(d(j)\) when the target \(t\) is fixed and clear from context.

\begin{lemma}\label{lem:k_values}
    Let \(Obj(k) = k e^{-\beta \sum_{j=1}^{k}d(j)}\) for \(k \in [s]\). We have that \(Obj(k + 1) > Obj(k)\) if and only if 
        \(k/(k+1) < e^{-\beta d(k+1)}\).
\end{lemma}
\begin{proof}
    When \(Obj(k + 1) > Obj(k)\), we have that \[ke^{-\beta \sum_{j=1}^{k}d(j)} < (k+1)e^{-\beta \sum_{j=1}^{k+1}d(j)}.\] This implies
    \(k/(k+1) < e^{-\beta d(k+1)}\).
\end{proof}

Since \(e^{-\beta d(k+1)}\) is decreasing and \(k/(k+1)\) is increasing for growing \(k\) we can use binary search to find the largest \(k\) for which this inequality is satisfied. 

The evaluation of \(k/(k+1) < e^{-\beta d(k+1)}\) requires the access to \(d(k+1)\) so the \((k+1)\)-th closest distance to the current target. In the next section, we describe how to efficiently retrieve the distance.

\subsection{Efficient distance retrieval for single target}
\label{appendix:single-target-distance}

The computation of the optimal \(k^*\) requires access to the \(k\)-th smallest distance to the target point \(t\). Moreover, evaluating the objective function for \(k^*\) requires the sum of the \(k\) smallest distances to the target point. In the following, we introduce a single-target data structure that supports these operations efficiently for a target point \(t\).

Let \(p = (p_0,\dots,p_{n-1})\) be a sorted array of integers with \(p_i \le p_{i+1}\) for all \(i\). For a target \(t \in \mathbb{Z}\), define the single-target distance 
\[d_t(x) := |x - t|.\]

The data structure \(S(p,t)\) supports: (i) retrieval of the \(k\)-th smallest distance, (ii) retrieval of the sum of the \(k\) smallest distances, and (iii) target updates.

\subsubsection{Initialization}
We partition \(p\) into two regions according to its relation to the value \(t\). Define the index
\[j = \min\{ i \mid p_i \ge t \}.\]
This index partitions \(p\) into two disjoint regions: elements strictly to the left of the target (\(p_i < t\)) and elements to the right or equal to the target (\(p_i \ge t\)).
Additionally, we compute the prefix-sum array \(a = (a_1, \dots, a_n)\) with \(a_i := \sum_{v=0}^{i-1} p_v\).

\subsubsection{Retrieving \(k\)-th closest Distance}
The KthDistance(\(k\)) operation selects \(k\) elements by choosing \(l\) elements from the left region and \(k-l\) elements from the right region.
The maximum selected distance is given by
\[h(l) := \max\bigl(t - p_{j-l},\ p_{j+(k-l)-1} - t\bigr).\]

The left term of \(h(l)\) increases with \(l\), while the right term decreases with \(l\). 
Thus, the minimizing value of \(h(l)\) (corresponding to the \(k\)-th smallest distance) can be found via a binary search over the number of elements \(l\) taken from the left.

\subsubsection{Sum of \(k\) Closest Distances}
The sum of the \(k\) smallest distances is computed using the same split point \(l\) obtained during the execution of KthDistance(\(k\)). By maintaining the prefix sum array \(a\), the sum of the \(k\)-th closest distances can be computed efficiently.

The \(l\) closest values to \(t\) from the left region are stored in the index interval \((j - l, j - 1)\). 
The sum of their distances to \(t\) is therefore
\[\sum_{i = j - l}^{j-1} (t - p_i) = l\cdot t - \sum_{i = j - l}^{j-1} p_i = l \cdot t - (a[j] - a[j -l]).\]

The remaining \(r = k - l\) elements are taken from the right region. 
These values are stored in the index interval \((j, j + r - 1)\). 
Their contribution to the distance sum is given by
\[\sum_{i = j}^{j + r - 1}(p_i - t) = \sum_{i = j}^{j + r - 1} p_i - r\cdot t = (a[j + r] - a[j]) - r \cdot t.\]

Combining both contributions yields the total sum of the \(k\) closest distances to \(t\):
\[l \cdot t - (a[j] - a[j -l]) + (a[j + k - l] - a[j]) - (k - l) \cdot t.\]

\subsubsection{Target Updates}
For an update of the target to a new value \(t'\), recomputing the index \(j\) via binary search restores all invariants.

\begin{lemma}
Initialization takes \(O(n)\) time. 
Each of KthDistance, SumKDistances, and UpdateTarget runs in \(O(\log n)\) time.
\end{lemma}
\begin{proof}
Initialization requires a scan of \(p\) to compute the prefix-sum array \(a\), giving a complexity of \(O(n)\).
For KthDistance, SumKDistances, and UpdateTarget, each operation requires a binary search that dominates the computational costs. Therefore, for each operation, the running time is bounded by \(O(\log n)\).
\end{proof}

We are now able to provide the running time for finding the optimal number of shifted values \(k^*\).
\begin{lemma}\label{lem:opt_k_time}
    Finding the optimal \(k^*\) for a fixed target \(t\) takes \(O(\log^2 d)\) time.
\end{lemma}
\begin{proof}
    The computation consists of two binary searches. The outer one is over the number of shifted values \(k\) as described in Lemma~\ref{lem:k_values}. The number of iteration is therefore bounded by \(O(\log d)\) where each iteration computes the \((k+1)\)-th closest distance in \(O(\log d)\) time with the single-target data structure \(S\).
\end{proof}

\subsection{Efficient computation of smooth-sensitivity}
We now present Algorithm \ref{alg:smooth_sens} to summarize the discussion in this section. We note that for better readability Algorithm \ref{alg:smooth_sens} only handles the case that the weight of the edge \(e \in E(v)\) is increased. For the case that \(w_e\) decreases by 1, the target value is set to \(\lambda - w_e\) instead of \(\lambda - w_e - 1\).
\begin{algorithm}
    \caption{\textsc{Efficient computation of smooth sensitivity}}\label{alg:smooth_sens}
    \begin{algorithmic}[1]
        \Statex \textbf{Input:} \(G=(V, E, w)\), \(v \in V\), \(\Delta_v\), \(\beta >0\), threshold \(\lambda\)
        \Statex \textbf{Output:} \(\beta\)-smooth sensitivity of \(f'_v\)
        \State \(s \gets 0\) \Comment{Initialization}
        \For{\(e \in E(v)\)}
            \State \(c \gets \{w_j + w'_{ij} : j \in \Delta_v(e) \}\)
            \State \(\mathcal{T} \gets c \cup \{\lambda - 1 - w_e\}\) \Comment{Target set}
            \State Let \(t_0 \leq \dots \leq t_d\) be the elements of \(\mathcal{T}\). 
            \State We initialize \(S(c,t_0)\) as the single target distance datastructure
            \For{\(\mathsf{i} \in \{0, \dots, d\}\)}
                \State \label{line:find-optimal} Find optimal \(z^*\) with \(k^*\) for target \(t_{\mathsf{i}}\). \Comment{Appendix \ref{appendix:smooth_props}}
                \State \label{line:find-opt-value} \(s \gets \max\left\{s, k^*e^{-\beta |z^*|}\right\}\). 
                \State If \(\mathsf{i} \neq d\), UpdateTarget(\(t_{i+1}\)).
            \EndFor
        \EndFor
        \State \Return \(s\)
    \end{algorithmic}
\end{algorithm}
We iterate over the targets \(t \in \mathcal{T}\) in nondecreasing order. For each target \(t\), we apply the algorithm from Appendix~\ref{appendix:smooth_props} to determine the optimal number of values \(k^*\) that should be connected to \(t\). If this yields a larger objective value, we update \(s\). Finally, as we move to the next target, we call UpdateTarget for the single-target data structure \(S\).

\begin{lemma} [Computation time of Algorithm \ref{alg:smooth_sens}]\label{lem:smooth_bias_run_time}
    Algorithm~\ref{alg:smooth_sens} computes the \(\beta\)-smooth sensitivity of \(f'_v\) using the biased estimator \(B'_T\) for a node \(v\) with degree \(d\) in \(O(d^2 \log^2 d)\) time.
\end{lemma}
\begin{proof}
    The node \(v\) has \(d\) incident edges and therefore the outer for loop is executed \(d\) times.
    Initializing the single-target data structure \(S(c,t_0)\) takes \(O(d)\) time. By Lemma~\ref{lem:opt_k_time}, Line~\ref{line:find-optimal} runs in \(O(\log^2 d)\) time and Line~\ref{line:find-opt-value} takes \(O(\log^2 d)\) by using SumKDistance for the computation of \(|z^*|\). Lastly, updating \(S\) takes \(O(\log d)\) time. The inner loop is executed \(O(d)\) where each iteration runs in \(O(\log^2 d)\). Hence, the total running time is in \(O(d^2 \log^2 d)\).
\end{proof}

\section{Efficient smooth sensitivity computation for unbiased estimator}
\label{appendix:smooth_unbiased}
We now propose the algorithm for the smooth sensitivity of \(f'_v(w^v)\) using the unbiased estimator \(U'_T\). It follows the same idea of shifting values to a target as in Appendix~\ref{appendix:smooth_bias} but requires a more detailed analysis since \(U'_T\) can take four distinct values compared to the two of \(B'_T\).
We recall the following notations. 
For a node \(v \in V\), let \(E(v) = \{ v \in e : e\in E \}\) denote the set of incident edges. For each \(i \in E(v)\), let \(\Delta_v(i) = \{j \in E(v) : \exists\{i,j,k\} \in \Delta_v\}\) denote the set of incident edge indices that form a triangle together with edge \(i\) and a non-incident edge of \(v\). 
Moreover, let \(u(i)\) and \(u(j)\) denote the other endpoints of edges \(i\) and \(j\), respectively.

In this section, we provide the smooth sensitivity algorithm from Section~\ref{sec:smooth} for the unbiased estimator \(U'_T\).
We first recall the reparameterized definition of the smooth sensitivity for the count of local below-threshold triangles \(f'_v\).
\[
    S^*_{f'_v,\beta}(w) = \max_{b \in B}\max_{z \in \mathbb{Z}^d}\left|f'_v(w +z + b)- f'_v(w + z)\right|e^{-\beta \cdot |z|}
\]
Here, \(b=(b_j)_{j\in E(v)}\) is a vector with exactly one nonzero entry, where \(b_j\in\{-1,0,1\}\).
It specifies which edge weight is increased or decreased by one.
Appendix~\ref{subsec:smooth_unb_alg} iterates over all \(2d\) possible choices of \(b\). We then consider the case when \(b\) is fixed. Suppose that \(i\) is the index of \(b\) such that \(b_i \neq 0\).

With these definitions, we expand the formulation of the smooth sensitivity explicitly for the unbiased estimator and a fixed \(i\):
\[
    \max_{z \in \mathbb{Z}^d}\left|\sum_{j \in \Delta_v(i)} U'_T(w_T + z_j + z_i + b_i) - U'_T(w_T + z_j + z_i)\right|e^{-\beta \cdot |z|}
\]
We introduce
\(y(j) = U'_T(w_T + z_j + z_i + b_i) - U'_T(w_T + z_j + z_i)\),
representing the contribution of \(j \in \Delta_v(i)\) to the sum.
This yields
\begin{equation}\label{eq:unb_smooth}
    \max_{z \in \mathbb{Z}^d}\left|\sum_{j \in \Delta_v(i)} y(j)\right|e^{-\beta \cdot |z|}
\end{equation}
We compute the maximum corresponding to a positive sum and a negative sum separately and take the maximum of the two in order to evaluate the absolute value:
\[
\max\left\{\max_{z \in \mathbb{Z}^d}\sum_{j \in \Delta_v(i)} y(j)e^{-\beta \cdot |z|}, \max_{z \in \mathbb{Z}^d}-\sum_{j \in \Delta_v(i)} y(j)e^{-\beta \cdot |z|}\right\}
\]
We refer to these as the objective functions with positive and negative sum contribution, respectively.
In the following we fix \(b_i = 1\), the same arguments and algorithm can be applied to the case \(b_i = -1\).

\subsection{Optimal solution for positive sum contribution}
In this section, we analyze the case where the sum contributes positively to the objective function:
\begin{equation}\label{eq:unb_smooth_pos}
    \max_{z \in \mathbb{Z}^d}\sum_{j \in \Delta_v(i)} y(j)e^{-\beta \cdot |z|}
\end{equation}

In the biased setting, the difference between the two estimators \(y(j)\) for each \(j \in \Delta_v(i)\) takes values in \(\{0,1\}\), corresponding to whether a triangle flips or remains unchanged.
In contrast, in the unbiased setting, \(y(j)\) may attain three distinct values.
For \(b_i = 1\), the possible values are
\(0\), \(\frac{p}{(1-p)^2}\), and \(-1-\frac{2p}{(1-p)^2}\).
We introduce the shorthand \(x = \frac{p}{(1-p)^2}\) to simplify notation. 
Let \(c = \left(w_j + w_{u(i)u(j)}\right)_{j \in \Delta_v(i)}\) denote the partial triangle weights, and define \(t_i = \lambda - 1 - w_i - z_i\) as the target value.
We partition the index set \(\Delta_v(i)\) into three disjoint subsets \(M, N, O\).
Specifically, \(M\) denotes the set of values adjacent to the target:
\[
    M = \left\{j \in \Delta_v(i): c_j = t_i-1 \lor c_j = t_i +1 \right\}
\]
The set \(N\) consists of values exactly on the target:
\[
    N = \left\{ j \in \Delta_v(i) : c_j = t_i \right\}.
\]
Finally, \(O\) contains all remaining values:
\[
    O = \left\{ j \in \Delta_v(i) : c_j < t_i - 1 \lor c_j > t_i + 1 \right\}.
\]
By construction, the sets \(M\), \(N\), and \(O\) form a partition of \(\Delta_v(i)\).

\begin{lemma}\label{lem:smooth_sum_con}
     For a solution \(z\) to optimization problem~\eqref{eq:unb_smooth_pos} with \(j \in \Delta_v(i)\) and \(z_j = 0\), the contribution \(y(j)\) satisfies
    \[
        y(j) = \begin{cases}
            x &\text{if } j \in M\\
            -1-2x &\text{if } j \in N\\
            0 &\text{if } j \in O
        \end{cases}
    \]
\end{lemma}
\begin{proof}
    Under the assumption \(z_j = 0\), the contribution of index \(j\) simplifies to
    \[
        y(j) = U'_T(c_j + w_i + z_i + 1) - U'_T(c_j + w_i + z_i).
    \]
    Consider first \(j \in M\) with \(c_j = t_i - 1\).
    In this case, we obtain
    \[
        U'_T(t_i - 1 + w_i + z_i + 1) - U'_T(t_i - 1 + w_i + z_i).
    \]
    Substituting the definition of \(t_i\) yields
    \[
        U'_T(\lambda - 1) - U'_T(\lambda - 2) = x.
    \]
    For \(j \in N\), the contribution evaluates to
    \(y(j) = U'_T(\lambda) - U'_T(\lambda - 1) = -1 - 2x\).
    Finally, for \(j \in O\), both
    \(U'_T(c_j + w_i + z_i + 1)\) and
    \(U'_T(c_j + w_i + z_i)\) are equal, implying \(y(j) = 0\).
\end{proof}

\subsection{Finding optimal shift sets}
We now determine how to set the values \(z_j\) for each of the sets \(M\), \(N\), and \(O\).
Let \(k_a\), for \(a \in \{M,N,O\}\), denote the number of values shifted from set \(a\).
We say that a value \(c_j\) is shifted if \(z_j > 0\).

\subsubsection{Shifting values from \(M\)}
We first show that no value in \(M\) is shifted.

\begin{lemma}\label{lem:m_opt}
    In any optimal solution \(z^*\), it holds that, for all \(j \in M\), \(z^*_j = 0\) and \(y(j) = x\).
\end{lemma}
\begin{proof}
    Let \(j \in M\) and consider \(z_j = 0\).
    By Lemma~\ref{lem:smooth_sum_con}, the contribution of \(j\) is \(y(j) = x\).
    Moreover, this is the maximum possible contribution of a single value.
    Increasing \(z_j\) strictly decreases the objective value due to the multiplicative factor \(e^{-\beta |z|}\).
    Hence, in any optimal solution, \(z_j = 0\) for all \(j \in M\) must hold.
\end{proof}

Consequently, it is always optimal to set \(z_j = 0\) for all \(j \in M\), and we have \(k_M = |M|\).

\subsubsection{Shifting values from \(N\)}
Next, we show that it is always optimal to shift all values in \(N\) before shifting any value from \(O\).

\begin{lemma}\label{lem:n_opt}
    Let \(z^*\) be an optimal solution.
    If there exists \(j \in N\) such that \(|z^*_j| \neq 1\), then, for all \(o \in O\), \(z^*_o = 0\).
\end{lemma}
\begin{proof}
    Let \(o \in O\) such that \(z^*_o \neq 0\), and assume there exists \(n \in N\) with \(|z^*_n| \neq 1\).
    We first consider the case \(z^*_n = 0\).
    Then, \(n\) contributes \(-1-2x\) to the sum, while \(o\) contributes \(x\), yielding
    \[
        \sum_{j\in\Delta_i(v)} y(j)
        =
        \sum_{j\in\Delta_i(v) \backslash \{o,n\} } y(j) + x - 1 - 2x.
    \]
    We construct a new solution \(z'\) by setting
    \(z'_j = z^*_j\) for all \(j \in \Delta_i(v)\setminus\{o,n\}\),
    \(z'_o = 0\), and \(z'_n = 1\).
    Clearly, \(|z'| \le |z^*|\). 
    Moreover, \(o\) now contributes \(0\), while \(n\) contributes \(x\), giving
    \[
        \sum_{j=1}^{s} y(j)
        =
         \sum_{j\in\Delta_i(v) \backslash \{o,n\} } y(j) + x,
    \]
    which strictly improves the objective value, contradicting the optimality of \(z^*\). 
    If instead \(|z^*_n| > 1\), then \(n\) contributes \(0\) to the sum.
    Decreasing \(|z^*_n|\) to \(1\) causes \(n\) to contribute \(x\), which again improves the objective value.
\end{proof}

Lemma~\ref{lem:n_opt} implies that all values in \(N\) must be shifted before any value in \(O\) is shifted.
We now determine the optimal number of values shifted from \(N\).

We begin with the solution \(z\) such that \(z_j = 0\) for all \(j \in \Delta_i(v)\).
In this configuration, values in \(M\) contribute \(x\), values in \(N\) contribute \(-1-2x\), and values in \(O\) contribute \(0\).
Shifting a value from \(N\) changes its contribution from \(-1-2x\) to \(x\).
Thus, the objective function as a function of \(k_N\) is
\[
    ((|M|+k_N)x + (|N|-k_N)(-1-2x))e^{-\beta (k_N + |z_i|)}
\]
Since \(|z|\) is fixed when optimizing over \(k_N\), the factor \(e^{-\beta |z_i|}\) is a positive constant independent of \(k_N\) and therefore does not affect the maximizer of the objective function.
Hence, rewriting this expression yields
\[
    obj(k_N)=\bigl(|M|x + (-1-2x)|N| + (1+3x)k_N\bigr)e^{-\beta k_N}.
\]
Let \(A = |M|x + (-1-2x)|N|\) and \(B = 1+3x\), and define
\(obj(k_N) = (A + Bk_N)e^{-\beta k_N}\).
Differentiating with respect to \(k_N\) gives
\[
    \frac{d}{dk_N} obj(k_N)
    =
    \bigl(B - \beta(A + Bk_N)\bigr)e^{-\beta k_N}.
\]
Since \(e^{-\beta k_N} > 0\) for all \(k_N\), the unique root satisfies
\(B - \beta(A + Bk_N) = 0\), yielding
\[
    k_N = \frac{1}{\beta} - \frac{A}{B}.
\]
Substituting \(A\) and \(B\) gives
\[
    k^*_N
    =
    \frac{1}{\beta}
    -
    \frac{|M|x + (-1-2x)|N|}{1+3x},
\]
which identifies the point beyond which shifting additional values from \(N\) is no longer beneficial.
Since \(k_N\) may be fractional, we select either \(\lfloor k_N \rfloor\) or \(\lceil k_N \rceil\), depending on which yields a higher objective value.
If \(k_N < |N|\), we found the optimal solution, otherwise, we proceed to shift values from \(O\).

\subsubsection{Shifting values from \(O\)}
We first show that any shifted value from \(O\) is always shifted to the closer of \(t_i - 1\) or \(t_i + 1\).

\begin{lemma}
    Let \(z^*\) be an optimal solution and let \(j \in O\) with \(z^*_j \neq 0\).
    Then
    \[
        |z^*_j|
        =
        \min\{ |c_j - t_i - 1|,\ |c_j - t_i + 1| \}.
    \]
\end{lemma}
\begin{proof}
    If \(z^*_j \neq t_i - 1 - c_j\) and \(z^*_j \neq t_i + 1 - c_j\), then \(y(j) \le 0\) implies that setting \(z_j = 0\) results in a strictly higher objective value.
    For both choices \(z^*_j = t_i - 1 - c_j\) and \(z^*_j = t_i + 1 - c_j\), the contribution is \(y(j) = x\).
    Among these, the choice minimizing \(|z^*_j|\) maximizes the objective, proofing the claim.
\end{proof}

We now describe how to determine the optimal number of values shifted from \(O\). We note that it is optimal to shift the values from \(O\) in increasing order of their distance to \(t_i - 1\) or \(t_i + 1\).
At this stage, the sets \(M\) and \(N\) are fixed, and all values in \(M \cup N\) contribute \(x\).
Shifting \(k_O\) values from \(O\) yields the objective value
\[
    obj(k_O) = x(k_O + |N| + |M|)e^{-\beta (|N| + \sum_j^{k_O} d_t(j))}
\]
where \(d_t(j)\) denotes the \(j\)-th closest distance from a value of \(O\) to \(t - 1\) or \(t + 1\) and we drop the constant factor of \(e^{-\beta|z_i|}\).

\begin{lemma}\label{lem:o_opt}
    Let \(k\) denote the number of values shifted from \(O\).
    The \((k+1)\)-st value is shifted if and only if
    \[
        \frac{k + |M| + |N|}{k + |M| + |N| + 1}
        <
        e^{-\beta d_t(k+1)}.
    \]
\end{lemma}
\begin{proof}
    The condition \(obj(k+1) > obj(k)\) is equivalent to
    \[
        x(k+1+|M|+|N|)
        e^{-\beta (\sum_{j=1}^{k+1} d_t(j))}
        >
        x(k+|M|+|N|)
        e^{-\beta (\sum_{j=1}^{k} d_t(j))},
    \]
    which simplifies to the stated inequality.
\end{proof}

Since \(|M|\) and \(|N|\) are fixed, the left-hand side of the inequality is monotonically decreasing in \(k\), while the right-hand side is monotonically non-increasing.
Thus, we can use binary search to find the largest value of \(k\) for which the inequality holds and shift \(k+1\) values from \(O\) in the optimal solution.

Finally, we show that it suffices to consider only a linear number of target values.

\begin{lemma}\label{lem:target_count}
    There exists an optimal solution \(z\) such that
    \(z_i = \lambda - w_i - c_j\),
    \(z_i = \lambda - 1 - w_i - c_j\), or
    \(z_i = \lambda - 2 - w_i - c_j\)
    for some \(j \in [s]\), or \(z_i = 0\).
    Equivalently, there exists an optimal solution with target \(t\) such that
    \(t = c_j\), \(t = c_j - 1\), or \(t = c_j + 1\) for some \(j \in [s]\),
    or \(t = \lambda - 1 - w_i\).
\end{lemma}
\begin{proof}
    Let \(z^*\) be an optimal solution that does not satisfy the stated property, and let \(t^*\) denote its target.
    By construction, \(|t^* - c_j| \ge 2\) for all \(j \in [s]\), implying that the sets \(M\) and \(N\) are empty.
    In this case, the same arguments as in Lemma~\ref{thm:target} apply here, and we are able to generate another optimal solution \(z'\) from \(z^*\) that satisfies the claim.
\end{proof}

The preceding lemma implies that it suffices to consider \(O(d)\) distinct values for \(z_i\).
Thus, we can iterate over all candidate values while maintaining an efficient overall algorithm.

\subsection{Optimal solution for negative sum contribution}
We now consider the case in which the sum contributes negatively to the objective function:
\begin{equation}\label{eq:unb_smooth_neg}
    \max_{z \in \mathbb{Z}^d}
    \sum_{j \in \Delta_v(i)} -y(j)e^{-\beta |z|}.
\end{equation}

We reuse the definitions of the sets \(M\), \(N\), and \(O\) introduced in the analysis of the positive contribution case.
By Lemma~\ref{lem:smooth_sum_con}, for a solution \(z\) with \(z_j = 0\) for all \(j \in \Delta_v(i)\), the contributions satisfy
\(-y(j) = -x\) for \(j \in M\),
\(-y(j) = 1 + 2x\) for \(j \in N\), and
\(-y(j) = 0\) for \(j \in O\).
Consequently, in the negative contribution case, the roles of the sets \(M\) and \(N\) are effectively reversed.

\begin{lemma}
    In any optimal solution \(z^*\), it holds that for all \(j \in N\), \(z^*_j = 0\) and \(-y(j) = 1 + 2x\).
\end{lemma}
\begin{proof}
    Let \(j \in N\).
    Setting \(z^*_j = 0\) yields the maximum possible contribution of \(1 + 2x\) to the sum.
    Any choice with \(z^*_j \neq 0\) decreases the objective value in~\eqref{eq:unb_smooth_neg} due to \(e^{-\beta \cdot |z|}\).
\end{proof}
An argument analogous to that of Lemma~\ref{lem:n_opt} applies to the sets \(M\) and \(O\) in this setting.
Specifically, values from \(M\) must be shifted before any value from \(O\) is shifted.
Let \(0 \le k_M \le |M|\) denote the number of values shifted from \(M\).
Values from \(N\) and shifted values of \(M\) contribute \(1 + 2x\), while the remaining unshifted values in \(M\) contribute \(-x\).
Shifting \(k_M\) values from \(M\) results in \(|z| = k_M\).
The objective function, without constant factor \(e^{-\beta|z_i|}\), as a function of \(k_M\) is therefore given by
\[
    obj(k_M)= \bigl((1+2x)(|N| + k_M) - x(|M| - k_M)\bigr)e^{-\beta k_M}.
\]
Maximizing this expression yields an optimal number of shifted values
\[
    k^*_M=\frac{1}{\beta}-\frac{(1+2x)|N| - x|M|}{1 + 3x}.
\]
Evaluating the objective function at the neighboring integer values of \(k^*_M\) identifies the optimal number of values shifted from \(M\).
Finally, we consider shifting values from \(O\).
Let \(k_O\) denote the number of values shifted from \(O\).
The objective function can be written as
\[
    obj(k_O)=(1+2x)(k_O + |N| + |M|)e^{-\beta (|M| + \sum_{j=1}^{k_O} d_t(j))},
\]
where \(d_t(j)\) denotes the \(j\)-th smallest distance from the target \(t\).
In contrast to the positive contribution case, where distances were measured relative to \(t-1\) and \(t+1\), distances here are taken with respect to \(t\).

Using the same argument as in Lemma~\ref{lem:o_opt}, we apply binary search to find the largest value \(k\) satisfying
\[
    \frac{k + |M| + |N|}{k + |M| + |N| + 1}<e^{-\beta d_t(k+1)}.
\]
The optimal solution then shifts \(k_O = k + 1\) values from \(O\).
Here, we are able to reuse the single-target data structure from Appendix~\ref{appendix:single-target-distance} to compute \(d_t(k+1)\).

\subsection{Efficient distance retrieval for double target}
\label{subsec:double-target-distance}
The computation of \(k^*_O\) requires access to the \(k\)-th smallest distance to \(t-1\) or \(t+1\) for problem~\eqref{eq:unb_smooth_pos}. 
Moreover, evaluating the objective function for \(k^*_O\) requires the sum of the \(k\) smallest distances to the respective target points.
In the following, we introduce a data structure that supports these operations efficiently for the two target points \(t-1\) and \(t+1\).

Let \(p = (p_0,\dots,p_{n-1})\) be a sorted array of integers with
\(p_i \le p_{i+1}\) for all \(i\). For a target \(t \in \mathbb{Z}\), define the
double-target distance
\[
    d_t(x) := \min\bigl(|x-(t-1)|,\ |x-(t+1)|\bigr).
\]

The double-target data structure \(D(p,t)\) supports: (i) retrieval of the \(k\)-th smallest distance, (ii) retrieval of the sum of the \(k\) smallest distances, and
(iii) target updates.

\subsubsection{Initialization}
We partition \(p\) into regions according to its relation to the values
\(t-1\), \(t\), and \(t+1\). Define the indices
\begin{align*}
\ell_0 &= \min\{ i \mid p_i \ge t-1 \}, &
\ell_1 &= \min\{ i \mid p_i > t-1 \}, \\
r_0 &= \min\{ i \mid p_i \ge t+1 \}, &
r_1 &= \min\{ i \mid p_i > t+1 \}.
\end{align*}
These indices partition \(p\) into five disjoint region: \(p_i < t-1\), \(p_i = t-1\), \(p_i = t\), \(p_i = t+1\), and \(p_i > t+1\).
Additionally, we compute the prefix-sum array \(a = (a_1, \dots, a_n)\) with \(a_i := \sum_{j=0}^{i-1} p_j\).

\subsubsection{Retrieving \(k\)-th closest Distance}
Let \(z := (\ell_1-\ell_0) + (r_1-r_0)\) define the number of points on \(t-1\) or \(t+1\), and let \(m := r_0 - \ell_1\) define the number of points on \(t\).
The KthDistance(\(k\)) operation proceeds in three phases:
\begin{enumerate}
  \item If \(k \le z\), return \(0\).
  \item Else if \(k \le z+m\), return \(1\).
  \item Otherwise, let \(k' := k-z-m\) and consider only indices
        \(i<\ell_0\) and \(i\ge r_1\).
\end{enumerate}
The remaining distances form two sorted sequences. 
We select \(k'\) elements by choosing \(l\) elements from the left region and \(k'-l\) from the right region. 
The maximum selected distance is
\[
h(l) := \max\bigl((t-1)-p_{\ell_0-l},\ p_{r_1+(k'-l)-1}-(t+1)\bigr).
\]

The left term of \(h(l)\) increases with \(l\), while the right term decreases
with \(l\).  
Thus, the minimizing value of \(h(l)\) (corresponding to the \(k\)-th smallest distance) can be found via binary search.

\subsubsection{Sum of \(k\) Closest Distances}
The sum of the \(k\) smallest distances is computed using the same split point \(l\) obtained during the execution of KthDistance(\(k\)).
Equivalently to the KthDistance function, let \(z := (\ell_1-\ell_0) + (r_1-r_0)\) define the number of points on \(t-1\) or \(t+1\), and let \(m :=  r_0 - \ell_1\) define the number of points on \(t\).
Elements at distance zero contribute nothing to the sum, while \(m\) elements at distance one contribute exactly their count.
The remaining contributions are computed using prefix sums.

Let \(l\) denote the number of selected elements from the left region.
The \(l\) closest values to \(t-1\) among those satisfying \(p_i < t-1\) are stored in the index interval \((\ell_0 - l, \ell_0 - 1)\).
The sum of their distances to \(t-1\) is therefore
\[
    \sum_{i = \ell_0 - l}^{\ell-1} \left((t-1) -  p_i\right) = l\cdot(t-1) - \sum_{i = \ell_0 - l}^{\ell-1} p_i = l \cdot(t-1) - (a[\ell_0] - a[\ell_0 -l]).
\]

The remaining \(r = k - l - m - z\) elements are taken from the right region.
The \(r\) closest values to \(t+1\) among those satisfying \(p_i > t+1\) are stored in the index interval \((r_1, r_1 + r - 1)\).
Their contribution to the distance sum is given by
\[
    \sum_{i = r_1}^{r_1 +r -1}\left(p_i - (t+1)\right) = \sum_{i = r_1}^{r_1 +r - 1}p_i - r\cdot(t+1) = (a[r_1 + r] - a[r_1]) -  r \cdot (t+1)
\]
Combining all contributions yields the sum of the \(k\) closest distances to either \(t-1\) or \(t+1\):
\[
    m + l \cdot (t-1) - (a[\ell_0] - a[\ell_0 -l]) + (a[r_1 + r] - a[r_1]) -  r \cdot (t+1).
\]

\subsubsection{Target Updates}
For an update of the target to \(t'\), recomputing the indices \(\ell_0,\ell_1,r_0,r_1\) via binary search restores all invariants.

\begin{lemma}
Initialization takes \(O(n)\) time. 
Each of KthDistance, SumKDistances, and UpdateTarget runs in \(O(\log n)\) time.
\end{lemma}
\begin{proof}
Initialization requires a scan of \(p\) to compute the prefix-sum array \(a\), giving a complexity of \(O(n)\).
For KthDistance, SumKDistances, and UpdateTarget each requires a binary search dominating the costs. Therefore, for each operation the running time is bounded by \(O(\log n)\).
\end{proof}

\subsection{Algorithm}\label{subsec:smooth_unb_alg}
In this subsection, we present Algorithm~\ref{alg:smooth_unb}, which computes the \(\beta\)-smooth sensitivity of \(f'_v\) for the unbiased estimator \(U'_T\).
The algorithm explicitly handles the case \(b_i = 1\); the case \(b_i = -1\) is treated analogously by using the initial target \(\lambda - w_e\) instead of \(\lambda - 1 - w_e\).

\begin{algorithm}
    \caption{\textsc{Calculation of smooth sensitivity with unbiased estimator}}\label{alg:smooth_unb}
    \begin{algorithmic}[1]
        \Statex \textbf{Input:} \(G=(V, E, w)\), \(v \in V\), \(\Delta_v\), \(\beta >0\)
        \Statex \textbf{Output:} \(\beta\)-smooth sensitivity of \(f'_v\)
        \State \(s \gets 0\) \Comment{Initialization}
        \For{\(e \in E(v)\)}
            \State \(c \gets \{w_j + w'_{ij} : j \in \Delta_v(e)\}\)
            \State \(\mathcal{T} \gets \{c_j + \delta :  c_j \in c, \delta \in \{-1,0,1\}\} \cup \{\lambda - 1 - w_e\}\)
            \Statex \Comment{Target Set}
            \State Let \(t_0 \leq \dots \leq t_d\) be the elements of \(\mathcal{T}\). 
            \Statex \underline{Solve Problem~\eqref{eq:unb_smooth_pos}}:
            \State  Initialize double target data structure \(D(c, t_0)\).
            \For{\(\mathsf{i} \in \{0, \dots, |\mathcal{T}|-1\}\)}
                \State Find optimal number of values shifted from \(N\).
                \State If \(k^*_N < |N|\), update \(s\) if necessary, advance target in \(D\) to \(t_{i+1}\), and continue.
                \State Find optimal \(k^*_O\) with binary search.
                \State Update global maximum \(s\) if necessary.
                \State Advance target in \(D\) to \(t_{i+1}\).
            \EndFor
            \Statex \underline{Solve Problem~\eqref{eq:unb_smooth_neg}}:
            \State Initialize single-target data structure \(S(c, t_0)\). 
             \For{\(\mathsf{i} \in \{0, \dots, |\mathcal{T}|-1\}\)}
                \State Find optimal number of values shifted from \(M\).
                \State If \(k^*_M < |M|\), update \(s\) if necessary, advance target in \(S\) to \(t_{i+1}\), and continue.
                \State Find optimal \(k^*_O\) with binary search.
                \State Update global maximum \(s\) if necessary.
                \State Advance target in \(S\) to \(t_{i+1}\).
            \EndFor
        \EndFor
        \State \Return \(s\)
    \end{algorithmic}
\end{algorithm}
For each incident edge \(e \in E(v)\), the algorithm computes the partial triangle weights \(c\) and constructs the set of candidate target values \(\mathcal{T}\) which is processed in sorted order.
For each target, the algorithm computes the optimal solution to the optimization problems \eqref{eq:unb_smooth_pos} and \eqref{eq:unb_smooth_neg}.
In the positive contribution case, a double-target distance data structure is used to determine the optimal number of shifted values from \(N\), and, if necessary, from \(O\) via binary search. 
The negative contribution case is handled analogously using a single-target distance data structure from Appendix~\ref{appendix:single-target-distance}, first optimizing over \(M\) and then over \(O\).
In both cases, targets are updated efficiently between iterations.
The algorithm returns the maximum objective value over all incident edges and candidate targets.
We now analyze the running time of the proposed algorithm by first bounding the cost of computing the optimal solution for a fixed incident edge and a fixed contribution sign.
\begin{lemma}\label{lem:time_smooth_pos}
    The optimal solutions to optimization problems~\eqref{eq:unb_smooth_pos} and~\eqref{eq:unb_smooth_neg} can be computed in \(O(d\log^2 d)\) time.
\end{lemma}
\begin{proof}
    We analyze the running time for a fixed incident edge and a fixed sign of the contribution.

    The computation begins by initializing the corresponding target-distance data structure, which requires \(O(d)\) time.
    The algorithm then iterates over \(O(d)\) candidate target values.
    In each iteration, the optimal number of shifted values from \(N\) (or \(M\) in the negative case) is computed in constant time.
    If necessary, the optimal number of shifted values from \(O\) is determined via a binary search over an interval of size \(O(d)\).
    Each step of the binary search performs a distance query on the data structure, which takes \(O(\log d)\) time.
    Consequently, determining the optimal number of shifted values from \(O\) requires \(O(\log^2 d)\) time per target.
    Finally, updating the target in the data structure between consecutive iterations requires \(O(\log d)\) time.
    Combining these bounds, each iteration costs \(O(\log^2 d)\) time, and over \(O(d)\) target values the total running time is \(O(d\log^2 d)\).
\end{proof}
Using this bound, we next derive the overall running time of Algorithm~\ref{alg:smooth_unb}.

\begin{lemma}[Computation time of Algorithm~\ref{alg:smooth_unb}]\label{lem:smooth_unb_run_time}
    Algorithm~\ref{alg:smooth_unb} computes the \(\beta\)-smooth sensitivity of \(f'_v\) using the unbiased estimator \(U'_T\) for a node \(v \in V\) of degree \(d\) in \(O(d^2 \log^2 d)\) time.
\end{lemma}
\begin{proof}
    Algorithm~\ref{alg:smooth_unb} iterates over all incident edges \(e \in E(v)\), of which there are at most \(d\).
    For each such edge, the algorithm performs an initialization step followed by the computation of both the positive and negative sum contributions.

    For a fixed edge \(e\), the initialization requires enumerating all triangles containing \(e\) and computing the corresponding partial triangle weights.
    Since \(|\Delta_v(e)| \le d\), this step requires \(O(d)\) time to collect the values and \(O(d \log d)\) time to sort them.

    By Lemma~\ref{lem:time_smooth_pos}, the computation of the optimal solution for the positive contribution and for the negative contribution each requires \(O(d\log^2 d)\) time.
    Thus, the total time per incident edge is \(O(d\log^2 d)\).

    For \(d\) incident edges, this results in an overall running time of \(O(d^2 \log^2 d)\).
\end{proof}

\begin{proof}[Proof of Theorem~\ref{thm:smooth_run_time}]
    Follows from Lemma~\ref{lem:smooth_bias_run_time} and Lemma~\ref{lem:smooth_unb_run_time}.
\end{proof}

\section{Experiments}\label{appendix:experiments}
In this section we give a detail description on how we constructed the graph ML-TELE-278 for the experiments of Section~\ref{sec:experiments} and provide additional results on a sparse graph.

\subsection{Details for the ML-TELE-278 graph construction}

\begin{table}
\caption{Experiment Dataset Overview}
\label{table:data}
\centering
\begin{tabular}{c|c|c}
    \textbf{Property} & \textbf{ML-Tele-278} & \textbf{GMWCS} \\\hline
    \(n\) & 278 & 1618\\
    \(m\) & 38503 & 1847\\
    \(|\Delta|\) & 3542276 & 132 \\
    \(\max w_T\) & 214 & -255 \\
    \(\min w_T\) & 0 & -522 \\
\end{tabular}
\end{table}

In 2014, Telecom Italia, in collaboration with several research institutions, organized the "Telecom Italia Big Data Challenge"\footnote{\href{http://theodi.fbk.eu/}{from BigDataChallenge contest}}, releasing an anonymized telecommunications dataset for the city of Milan \cite{barlacchi2015multi}. The dataset ("Telecommunications - MI to MI") records aggregated communication intensity between geographic areas of Milan, which are partitioned into a 100×100 grid of equally sized cells. For every pair of areas, it reports the volume of telecommunication activity in 10-minute intervals.

In our analysis, we consider the 10-minute interval on Sunday, November 3, 2013, from 19:50 to 20:00 (local time). We assume that each base station serves a 6×6 block of cells, which yields a total of 278 base stations across the city. From this, we construct an undirected weighted graph in which each node corresponds to a radio base station, each edge represents a connection between two stations, and each edge weight is the number of calls between them.
Under local weight differential privacy, the privacy of individual calls is protected. In particular, publishing the number of below-threshold triangles does not reveal whether a specific individual made a call, nor where that call took place.

Let \(I_e\) denote the interaction intensity associated with an edge \(e \in E\) obtained from the dataset, and let \(L\) be the total number of calls that occurred during the selected 10-minute interval. We set
    \[w_e \;=\; L \cdot \frac{I_e}{\sum_{e' \in E} I_{e'}}\]
for all \(e \in E.\)
The total number of calls \(L\) is not explicitly reported in the challenge data. Assuming an average of two calls per person per day~\cite{kang2012towards} and a population of approximately 1.3 million in Milan~\cite{barlacchi2015multi}, we estimate \(L \approx 20{,}000\) for a 10-minute interval.
This estimate indicates that the number of edges \(m\) is much larger than \(L\), so most edges obtain weight zero. We summarize the resulting graph, which we refer to as \textsc{ML-Tele-278}, in Table~\ref{table:data}.

\subsection{Experiment on GMWCS}

We next evaluate our methods on a second graph, denoted \textsc{GMWCS}. We use the largest instance from the benchmark set of \cite{loboda2016solving}, which was also included in the \emph{Generalized Maximum-Weight Connected Subgraph} (GMWCS) track of the \emph{11th DIMACS Implementation Challenge in Collaboration with ICERM: Steiner Tree Problems}. The properties of this instance are summarized in Table~\ref{table:data}.
Although this graph has more nodes than \textsc{ML-Tele-278}, it is extremely sparse. The average degree is only slightly above two, and it contains only 132 triangles. This setting is difficult for our algorithm: each node contributes noise, but the true below-threshold triangle count is very small. Moreover, Figure~\ref{fig:tele_all} already suggests that our approach performs less favorably on graphs with very few triangles.
Nevertheless, Figure~\ref{fig:gmwcs_lambda} shows that we can still release meaningful statistics. In particular, the biased estimator with smooth sensitivity achieves an average relative error that is not much larger than one. While the non-interactive baseline attains lower error than all variants of our algorithm, the gap is typically within a factor of five.

\begin{figure}
    \centering
    \begin{tikzpicture}[scale=0.5]
        \begin{axis}[
            xlabel={\(\lambda\)},
            ylabel={relative error},
            grid=both,
            grid style={dotted,gray!30},
            ymode=log,
            legend pos=outer north east, 
            legend style={
                draw=none,                
                cells={anchor=west},
                font=\Large,
                row sep=6pt              
            },
            tick label style={font=\small},
        ]

        \addplot+[thick, color=blue] table [x=x, y=smooth_unbiased_l2_rel, col sep=comma] {fig/gmwcs_lambda_rel_error.csv};
        \addlegendentry{smooth-unbiased}

        \addplot+[thick, color=red] table [x=x, y=naive_l2_rel, col sep=comma] {fig/gmwcs_lambda_rel_error.csv};
        \addlegendentry{baseline}

        \addplot+[mark=diamond, thick, color=purple] table [x=x, y=smooth_biased_l2_rel, col sep=comma] {fig/gmwcs_lambda_rel_error.csv};
        \addlegendentry{smooth-biased}

        \addplot+[mark=o, thick, color=orange] table [x=x, y=global_unbiased_l2_rel, col sep=comma] {fig/gmwcs_lambda_rel_error.csv};
        \addlegendentry{global-unbiased}

        \addplot+[mark=square, thick, color=green] table [x=x, y=global_biased_l2_rel, col sep=comma] {fig/gmwcs_lambda_rel_error.csv};
        \addlegendentry{global-biased}

        \end{axis}
    \end{tikzpicture}
    \caption{Relative error for varying \(\lambda\) on the input graph GMWCS}
    \label{fig:gmwcs_lambda}
\end{figure}

\subsection{Experiments on assignment functions}\label{subsec:exp_assign}
In this section, we present empirical results to compare the assignment strategies we proposed in Section~\ref{sec:mincov}.

\definecolor{mcmfcolor}{RGB}{31, 119, 180}
\definecolor{greedycolor}{RGB}{255, 127, 14}
\definecolor{randomcolor}{RGB}{214, 39, 40}
\definecolor{degcolor}{RGB}{44, 160, 44}

\pgfplotsset{
    benchmark style/.style={
        width=0.9\linewidth,
        height=6cm,
        xlabel={Number of Nodes \(n\)},
        grid=major,
        grid style={dashed, gray!30},
        legend style={
            at={(0.02,0.98)},
            anchor=north west,
            font=\small,
            draw=gray!50
        },
        every axis plot/.append style={line width=1.2pt},
        tick label style={font=\small},
        label style={font=\small},
        xticklabel style={font=\small},
    }
}

\pgfplotstableread[col sep=comma]{fig/complete.csv}\completedata
\begin{figure}[h]
    \centering
    \begin{subfigure}[b]{0.48\textwidth}
        \centering
        \begin{tikzpicture}[scale=0.7]
        \begin{axis}[
            benchmark style,
            ylabel={Running Time (s)},
            xlabel={Number of Triangles \(|\Delta|\)},
            legend to name=sharedlegendcomplete,
            legend columns=4,
            xtick={1000, 5000, 10000, 15000},
        ]
        \addplot+[color=mcmfcolor, mark=o]
            table[x=triangle_count_mean, y=mcmf_time_mean] {\completedata};
        \addplot+[color=greedycolor, mark=square]
            table[x=triangle_count_mean, y=greedy_time_mean] {\completedata};
        \addplot+[color=randomcolor, mark=triangle]
            table[x=triangle_count_mean, y=random_time_mean] {\completedata};
        \addplot+[color=degcolor, mark=diamond]
            table[x=triangle_count_mean, y=deg_time_mean] {\completedata};
        \legend{MCMF, Greedy, Random, Degeneracy}
        \end{axis}
        \end{tikzpicture}
        \caption{Running time}
        \label{fig:complete_time}
    \end{subfigure}%
    \hspace{0.01\textwidth}%
    \begin{subfigure}[b]{0.48\textwidth}
        \centering
        \begin{tikzpicture}[scale=0.7]
        \begin{axis}[
            benchmark style,
            ylabel={Assignment Cost \(\sum \binom{\ell_e}{2}\)},
            xlabel={Number of Triangles \(|\Delta|\)},
            xtick={1000, 5000, 10000, 15000},
        ]
        \addplot+[color=mcmfcolor, mark=o]
            table[x=triangle_count_mean, y=mcmf_cost_mean] {\completedata};
        \addplot+[color=greedycolor, mark=square]
            table[x=triangle_count_mean, y=greedy_cost_mean] {\completedata};
        \addplot+[color=randomcolor, mark=triangle]
            table[x=triangle_count_mean, y=random_cost_mean] {\completedata};
        \addplot+[color=degcolor, mark=diamond]
            table[x=triangle_count_mean, y=deg_cost_mean] {\completedata};
        \end{axis}
        \end{tikzpicture}
        \caption{Assignment cost}
        \label{fig:complete_cost}
    \end{subfigure}%
    \hspace{0.01\textwidth}%
    \begin{subfigure}[b]{0.5\textwidth}
        \centering
        \begin{tikzpicture}[scale=0.7]
        \begin{axis}[
            benchmark style,
            ylabel={\(ALG / OPT - 1\)},
            xlabel={Number of Triangles \(|\Delta|\)},
            xtick={1000, 5000, 10000, 15000},
            ymode=log,
            xmin=10,
        ]
        \addplot+[color=greedycolor, mark=square]
            table[x=triangle_count_mean, y expr=\thisrow{greedy_rel_error_mean}] {\completedata};
        \addplot+[color=randomcolor, mark=triangle]
            table[x=triangle_count_mean, y expr=\thisrow{random_rel_error_mean}] {\completedata};
        \addplot+[color=degcolor, mark=diamond]
            table[x=triangle_count_mean, y expr=\thisrow{deg_rel_error_mean}] {\completedata};
        \end{axis}
        \end{tikzpicture}
        \caption{Relative error in assignment cost}
        \label{fig:complete_rel_cost}
    \end{subfigure}
    \vspace{0.3em}
    \ref{sharedlegendcomplete}
    \caption{Experimental results of assignment algorithms on complete graphs}
    \label{fig:complete}
\end{figure}

\pgfplotstableread[col sep=comma]{fig/delaunay.csv}\delaunaydata

\begin{figure}[t]
    \centering
    \begin{subfigure}[b]{0.49\textwidth}
        \centering
        \begin{tikzpicture}[scale=0.7]
        \begin{axis}[
            benchmark style,
            ylabel={Running Time (s)},
            xlabel={Number of Triangles \(|\Delta|\)},
            legend to name=sharedlegenddelaunay,
            xtick={1000, 2000, 3000, 4000},
            legend columns=4,
        ]
        \addplot+[color=mcmfcolor, mark=o]
            table[x=triangle_count_mean, y=mcmf_time_mean] {\delaunaydata};
        \addplot+[color=greedycolor, mark=square]
            table[x=triangle_count_mean, y=greedy_time_mean] {\delaunaydata};
        \addplot+[color=randomcolor, mark=triangle]
            table[x=triangle_count_mean, y=random_time_mean] {\delaunaydata};
        \addplot+[color=degcolor, mark=diamond]
            table[x=triangle_count_mean, y=deg_time_mean] {\delaunaydata};
        \legend{MCMF, Greedy, Random, Degeneracy}
        \end{axis}
        \end{tikzpicture}
        \caption{Running time}
        \label{fig:delaunay_time}
    \end{subfigure}
    \begin{subfigure}[b]{0.49\textwidth}
        \centering
        \begin{tikzpicture}[scale=0.7]
        \begin{axis}[
            benchmark style,
            ylabel={Assignment Cost \(\sum \binom{\ell_e}{2}\)},
            xlabel={Number of Triangles \(|\Delta|\)},
            xtick={1000, 2000, 3000, 4000},
        ]
        \addplot+[color=mcmfcolor, mark=o]
            table[x=triangle_count_mean, y=mcmf_cost_mean] {\delaunaydata};
        \addplot+[color=greedycolor, mark=square]
            table[x=triangle_count_mean, y=greedy_cost_mean] {\delaunaydata};
        \addplot+[color=randomcolor, mark=triangle]
            table[x=triangle_count_mean, y=random_cost_mean] {\delaunaydata};
        \addplot+[color=degcolor, mark=diamond]
            table[x=triangle_count_mean, y=deg_cost_mean] {\delaunaydata};
        \end{axis}
        \end{tikzpicture}
        \caption{Assignment cost}
        \label{fig:delaunay_cost}
    \end{subfigure}

    \vspace{0.3em}
    \ref{sharedlegenddelaunay}
    \caption{Experimental results of assignment algorithms on graphs generated via Delaunay triangulation.}
    \label{fig:delaunay}
\end{figure}

\paragraph{Experimental Setup.}
We evaluate the assignment cost and running time of four strategies: 
\emph{Min-Cost Max-Flow} (\emph{MCMF}), \emph{Greedy}, \emph{Degeneracy}, and a \emph{Random} baseline.
The \emph{Random} baseline assigns each triangle independently and uniformly at random to one of its three edges. The \emph{MCMF} algorithm follows the graph construction and the proof of Theorem \ref{thm:assgn_time}. 
The \emph{Degeneracy} algorithm follows the construction in the proof of Lemma~\ref{lem:deg_ass}, using the linear-time \(O(n + m)\) algorithm of~\cite{matula1983smallest} to compute the degeneracy ordering.
We evaluate all algorithms on three graph families: complete graphs \(K_n\), planar graphs generated via Delaunay triangulation, and subgraphs of a real-world Google web graph~\cite{leskovec2009community}, which was released in 2002 by Google as part of a programming contest. The web graph is part of the SNAP network datasets \cite{snapnets}.
Since the \emph{Greedy} and \emph{Random} strategies depend on the order in which triangles are processed, we execute each algorithm 20 times for planar and complete graphs and 5 times on the web graph on independently and uniformly shuffled triangle lists and report the mean.
Across all graph types and sizes, the standard deviation of both running time and assignment cost over these trials remained negligibly small, confirming that the chosen repetitions are sufficient to obtain stable estimates.
All experiments were conducted on a 2024 MacBook Pro with an Apple M4 Pro chip and 48 GB of memory.

\paragraph{Complete Graphs.}
Figure~\ref{fig:complete} reports assignment cost, running time, and relative cost with respect to \emph{MCMF} on complete graphs \(K_n\).
\emph{Greedy}, \emph{Random}, and \emph{Degeneracy} all run in under \(0.02\)\,seconds across all tested sizes, whereas \emph{MCMF} already exceeds \(5\)\,seconds at \(n = 40\) nodes (\(|\Delta| = 9{,}880\) triangles), confirming that computing the optimal solution is intractable at scale.
Figure~\ref{fig:complete_rel_cost} shows the relative error of the assignment cost, defined as \(ALG/OPT - 1\), where \(ALG\) denotes the cost obtained by each algorithm and \(OPT\) denotes the optimal cost obtained by \emph{MCMF}. \emph{Greedy} achieves roughly one order of magnitude lower relative error than \emph{Random} across all sizes, with gaps of approximately \(0.15\%\) versus \(5.5\%\) at \(n=40\).
The \emph{Degeneracy} ordering performs poorly on complete graphs. 
Nodes that appear early in the ordering are assigned edges connecting them to almost all other nodes, resulting in highly skewed loads. In this unstructured setting, both \emph{Greedy} and \emph{Random} distribute load far more evenly and therefore achieve more competitive costs.

\paragraph{Planar Graphs via Delaunay Triangulation.}
Figure~\ref{fig:delaunay} reports results on planar graphs generated by Delaunay triangulation.
As established in Lemma~\ref{lem:planar_opt}, an optimal assignment of cost zero always exists for planar graphs, and therefore \emph{MCMF} achieves it in all instances.
The running time profile mirrors that of complete graphs: \emph{MCMF} scales poorly while the three heuristics remain fast, with \emph{MCMF} reaching \(1.1\)\,seconds at \(|\Delta| \approx 3{,}200\) triangles compared to under \(0.015\)\,seconds for all heuristics.
In terms of assignment cost, the structural ordering of \emph{Degeneracy} now proves beneficial. It outperforms \emph{Random} by a factor of approximately \(3\times\) across all sizes, as the degeneracy ordering aligns well with the local triangle structure of planar graphs.
\emph{Greedy} retains the best cost among all non-optimal strategies, achieving costs roughly \(6\times\) lower than \emph{Random} and \(2\times\) lower than \emph{Degeneracy} at \(|\Delta| \approx 3{,}200\).

\paragraph{Google Web Graph.}
Figure~\ref{fig:degeneracy} reports results on subgraphs of the Google web graph, a real-world instance with bounded degeneracy of 44.
\emph{MCMF} is omitted as it is no longer tractable at this scale.
To preserve the power-law degree distribution characteristic of web graphs, we subsample by retaining each edge independently with probability \(p \in \{0.05, 0.1, 0.25, 0.5, 0.7, 0.8, 0.9, 1\}\), rather than sampling nodes. 
A node of degree \(d\) retains expected degree \(pd\) under edge sampling, preserving the relative ordering of degrees and thus the shape of the degree distribution.
For every value of \(p\), we generated 5 random subgraphs and we averaged the resulting costs and running time of all algorithms. 
The results reveal a pronounced and consistent separation between strategies as the graph density increases.
At a sampling fraction of \(p = 0.2\) (\(|\Delta| \approx 107{,}380\)), \emph{Greedy} achieves a cost of \(969\), compared to \(6{,}519\) for \emph{Degeneracy} and \(23{,}342\) for \emph{Random}.
At the largest fraction \(p = 1.0\), corresponding to over \(13\) million triangles, \emph{Greedy} achieves a cost of \(26{,}427{,}913\) versus \(41{,}021{,}836\) for \emph{Degeneracy} (\(1.6\times\) higher) and \(69{,}055{,}393\) for \emph{Random} (\(2.6\times\) higher).
Running times at this scale are \(61\)\,seconds for \emph{Greedy}, \(59\)\,seconds for \emph{Degeneracy}, and \(47\)\,seconds for \emph{Random}, all within the same order of magnitude.

\pgfplotstableread[col sep=comma]{fig/degeneracy.csv}\degeneracydata

\begin{figure}[t]
    \centering
    \begin{subfigure}[b]{0.49\textwidth}
        \centering
        \begin{tikzpicture}[scale=0.7]
        \begin{axis}[
            benchmark style,
            ylabel={Running Time (s)},
            xlabel={Number of Triangles \(|\Delta|\)},
            legend to name=sharedlegendweb,
            xtick={1000000, 4000000, 7000000, 10000000, 13000000},
            legend columns=4,
        ]
        \addplot+[color=greedycolor, mark=square]
            table[x=triangle_count_mean, y=greedy_time_mean] {\degeneracydata};
        \addplot+[color=randomcolor, mark=triangle]
            table[x=triangle_count_mean, y=random_time_mean] {\degeneracydata};
        \addplot+[color=degcolor, mark=diamond]
            table[x=triangle_count_mean, y=deg_time_mean] {\degeneracydata};
        \legend{Greedy, Random, Degeneracy}
        \end{axis}
        \end{tikzpicture}
        \caption{Running time.}
        \label{fig:degeneracy_time}
    \end{subfigure}
    \begin{subfigure}[b]{0.49\textwidth}
        \centering
        \begin{tikzpicture}[scale=0.7]
        \begin{axis}[
            benchmark style,
            ylabel={Assignment Cost \(\sum \binom{\ell_e}{2}\)},
            xlabel={Number of Triangles \(|\Delta|\)},
            xtick={1000000, 4000000, 7000000, 10000000, 13000000},
        ]
        \addplot+[color=greedycolor, mark=square]
            table[x=triangle_count_mean, y=greedy_cost_mean] {\degeneracydata};
        \addplot+[color=randomcolor, mark=triangle]
            table[x=triangle_count_mean, y=random_cost_mean] {\degeneracydata};
        \addplot+[color=degcolor, mark=diamond]
            table[x=triangle_count_mean, y=deg_cost_mean] {\degeneracydata};
        \end{axis}
        \end{tikzpicture}
        \caption{Assignment cost.}
        \label{fig:degeneracy_cost}
    \end{subfigure}

    \vspace{0.3em}
    \ref{sharedlegendweb}
    \caption{Experimental results of assignment algorithms on graphs sub-sampled from a Google web graph.}
    \label{fig:degeneracy}
\end{figure}


\end{document}